\documentclass[sigconf]{acmart}

\usepackage{amsmath,amsfonts}
\usepackage{algorithm,algorithmic}
\usepackage{graphicx}
\usepackage{textcomp}
\usepackage{xcolor}
\usepackage{subfigure}
\usepackage{xspace}
\usepackage{tikz}
\usepackage{xcolor}
\usepackage{makecell}
\usepackage{bm}
\usepackage{balance}

\newcommand{\camera}[1]{{\color{black}{#1}}}

\newcommand{\ouralg}{$\mathsf{PEMA}$\xspace}
\newcommand{\pone}{$\mathsf{ORA}$\xspace}
\newcommand{\opt}{$\mathsf{OPTM}$\xspace}
\newcommand{\ruleA}{$\mathsf{RULE}$\xspace}

\newcommand{\trainT}{\texttt{TrainTicket}\xspace}
\newcommand{\hotelR}{\texttt{HotelReservation}\xspace}
\newcommand{\socS}{\texttt{SockShop}\xspace}

\AtBeginDocument{%
  }

\copyrightyear{2022}
\acmYear{2022}
\setcopyright{rightsretained}

\acmConference[HPDC '22]{Proceedings of the 31st International Symposium on High-Performance Parallel and Distributed Computing}{June 27-July 1, 2022}{Minneapolis, MN, USA}
\acmBooktitle{Proceedings of the 31st Int'l Symposium on High-Performance Parallel and Distributed Computing (HPDC '22), June 27-July 1, 2022, Minneapolis, MN, USA}
\acmDOI{10.1145/3502181.3531460}
\acmISBN{978-1-4503-9199-3/22/06}

\usepackage{etoolbox}
\makeatletter
\patchcmd{\maketitle}{\@copyrightpermission}{
   \begin{minipage}{0.3\columnwidth}
     \href{https://creativecommons.org/licenses/by/4.0/}{\includegraphics[width=0.90\textwidth]{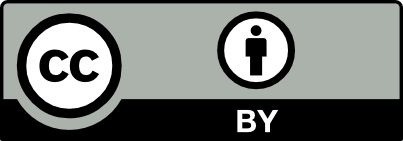}}
   \end{minipage}\hfill
   \begin{minipage}{0.7\columnwidth}
     \href{https://creativecommons.org/licenses/by/4.0/}{This work is licensed under a Creative Commons Attribution International 4.0 License.}
   \end{minipage}

   \vspace{5pt}
}{}{}
\makeatother

\begin{document}

\title{Practical Efficient Microservice Autoscaling with QoS Assurance}

\author{Md Rajib Hossen}
\affiliation{%
  \institution{The University of Texas at Arlington}
  \state{Texas}
  \country{USA}
  \postcode{76019}
}
\email{mdrajib.hossen@mavs.uta.edu}

\author{Mohammad A. Islam}
\affiliation{%
  \institution{The University of Texas at Arlington}
  \state{Texas}
  \country{USA}
}
\email{mislam@uta.edu}
  
  \author{Kishwar Ahmed}
\affiliation{%
  \institution{University of South Carolina Beaufort}
  \state{South Carolina}
  \country{USA}
}
\email{ahmedk@uscb.edu}


\begin{abstract}
Cloud applications are increasingly moving away from monolithic services to agile microservices-based deployments. However, efficient resource management for microservices poses a significant hurdle due to the sheer number of loosely coupled and interacting components. The interdependencies between various microservices make existing cloud resource autoscaling techniques ineffective. Meanwhile, machine learning (ML) based approaches that try to capture the complex relationships in microservices require extensive training data and cause intentional SLO violations. Moreover, these ML-heavy approaches are slow in adapting to dynamically changing microservice operating environments. In this paper, we propose \ouralg (\textbf{P}ractical \textbf{E}fficient \textbf{M}icroservice \textbf{A}utoscaling), a lightweight microservice resource manager that finds efficient resource allocation through opportunistic resource reduction. \ouralg's lightweight design enables novel workload-aware and adaptive resource management. Using three prototype microservice implementations, we show that \ouralg can find efficient resource allocation and save up to 33\% resource compared to the commercial rule-based resource allocations.
\end{abstract}

\ccsdesc[500]{Computer systems organization~Cloud computing}

\keywords{Autoscaling, microservices, resource management, cloud computing, quality of service}


\fancyhead{}
\maketitle

\section{Introduction}

\begin{figure}[t]
\includegraphics[width=0.45\textwidth]{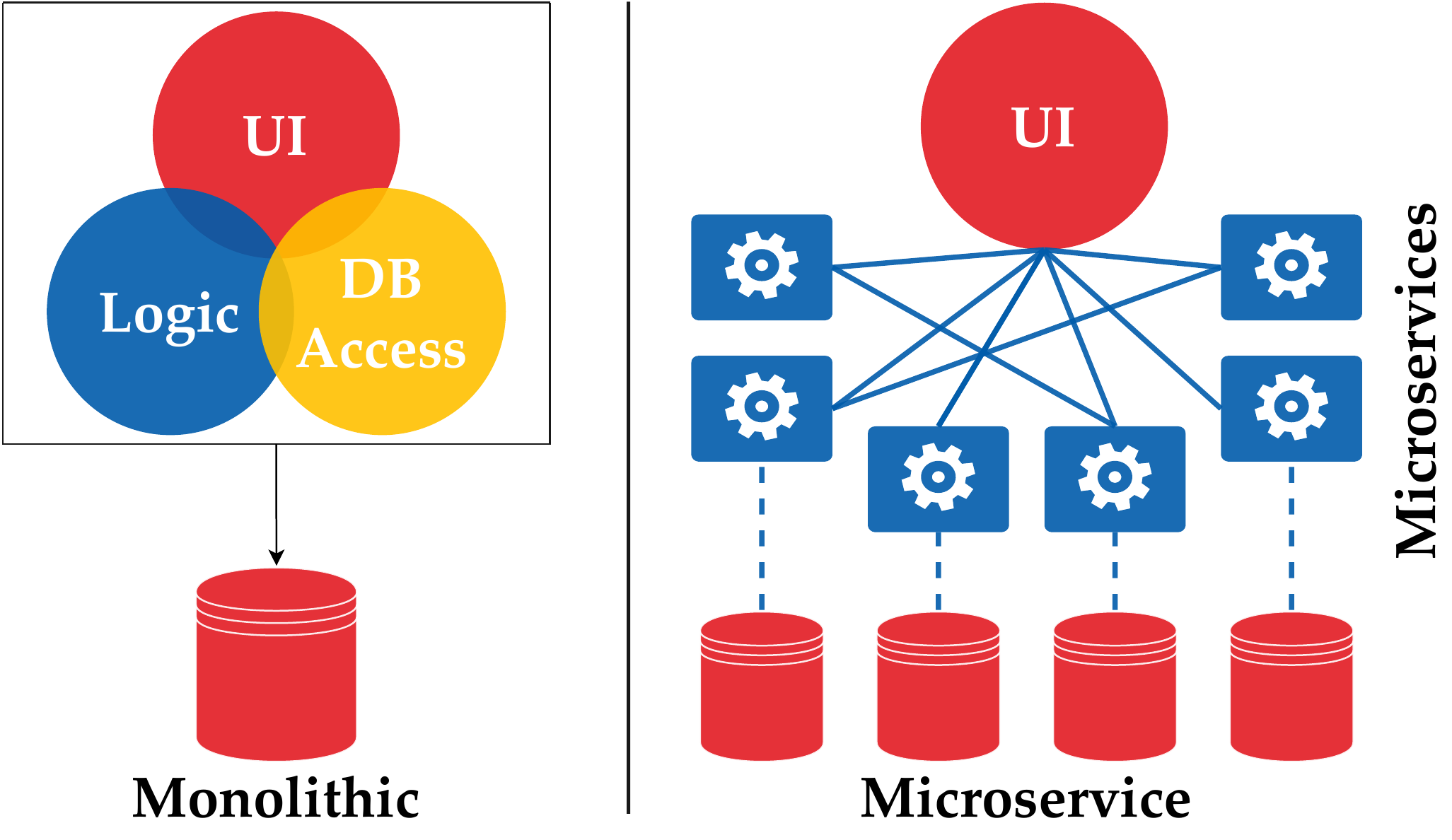}
\caption{Comparison between monolithic and microservices architecture. Microservices consists of many small loosely coupled systems.}
\label{fig:mono_vs_micro}
\end{figure}

\noindent\textbf{Motivation.}
Microservices architecture is enjoying a growing penetration in user-facing cloud applications where an ensemble of loosely-coupled and small service components (i.e., microservices) work together to serve user requests \cite{gan2018architectural,deathstarbench,heinrich2017performance}. As illustrated in Fig.~\ref{fig:mono_vs_micro}, microservices architecture is a significant departure from traditional monolithic deployments with a few large application layers such as user-facing front-end, back-end business logic, and database \cite{hou2021alphar}. Unlike monolithic applications, the small microservices can be easily managed and kept updated by small dedicated DevOps teams \cite{wolff2016microservices}. Moreover, microservices are typically stateless and communicate using lightweight APIs \cite{microservice_definiton, nginx_microservice}. Hence, they offer agile resource management and scaling, better fault tolerance, and great platform agnostic compatibility among different microservices that cannot be matched by monolithic applications \cite{wolff2016microservices,zhou2018overload}. 

Microservices come with their own sets of challenges, and in this paper, we focus on its resource management. In principle, microservice resource management is same as monolithic applications - achieve the desired performance (e.g., end-to-end response latency) with the minimum resource allocation \cite{gandhi2012autoscale,google_autoscale,azure_autoscale}. Resource management for microservices-based applications, however, is more challenging because these applications have a much larger configuration space due to the sheer number of microservices responsible for the application performance. For example, if we consider an application with $m$ microservices where each microservice can be configured with $n$ different CPU allocations, there will be $n^m$ possible resource configurations.
Moreover, microservices have complex communication topology and inter-dependencies that make it harder to identify and mitigate Quality of Service (QoS) violations \cite{deathstarbench,zhou2018overload}. A single user request may traverse through several microservices, and if any microservice in the critical path becomes a bottleneck, the end-to-end response time will increase significantly \cite{firm}.
Our motivating experiments on three prototype microservices show that the same amount of CPU allocation can result in more than 250\% increase in application latency based on how the resource is distributed among different microservices. Meanwhile, existing resource management techniques developed for monolithic applications with a few service layers cannot readily capture the complex microservice interactions to make effective resource allocation choices \cite{delimitrou2013paragon,delimitrou2014quasar,lo2014towards,lo2015heracles}. Nevertheless, addressing these resource management challenges for microservices is of paramount importance as an increasing number of production cloud services have been adopting microservice architectures \cite{zhang2021sinan,hou2021alphar}.

\noindent\textbf{Limitation of state-of-art approaches.} Owing to the growing interest, several recent works try to address the resource management challenges in microservices \cite{zhang2021sinan,firm,gan2021sage,gan2019seer,hou2021alphar}. They focus on utilizing machine learning (ML) techniques to capture the complex relationship between microservice resources and performance. 
For instance, FIRM \cite{firm} uses a combination of support vector machines (SVM) and reinforcement learning to localize root causes of SLO violations, and apply resource autoscaling to avert these violations. Sage \cite{gan2021sage}, on the other hand, uses supervised training to identify dependencies between different microservices using a Causal Bayesian Network, and a graph encoder to track the QoS violating microservices to adjust their resources. However, this line of works built on ML are fundamentally limited by their extensive training requirements, both in terms of training time to capture the dynamics of the microservices and data resolution (e.g., request level traces to build dependency graphs). More importantly, to learn from the data, some ML-based techniques intentionally cause or allow SLO violations which is undesirable in production systems \cite{firm,zhang2021sinan,gan2019seer,gan2021sage}.
Also, any changes in the microservices architecture and inter-dependencies will require retraining the system. This ML retraining can become a barrier for real world microservices applications which go through frequent software/code updates. 
ML retraining can also be triggered by changes in underlying cloud hardware due to server migrations and upgrades.
On the other hand, the resource demand of microservices changes with the workload on a daily basis. However, existing approaches focusing on SLO violations do not directly incorporate dynamic workload in their learning \cite{firm,zhang2021sinan,gan2019seer,gan2021sage}.

\noindent\textbf{Key insights and contributions.} 
To avoid the hurdles of the approaches mentioned above, we propose \ouralg (\textbf{P}ractical \textbf{E}fficient \textbf{M}icroservice \textbf{A}utoscaling), a lightweight microservice resource manager that does not need extensive training. \ouralg utilizes iterative feedback-based tuning to find efficient resource allocations that satisfy the SLO. Instead of finding the best resource configuration, \ouralg first allocates abundant resources to all microservices to satisfy SLO and then tries to exploit resource reduction opportunities.
\camera{Allocating abundant resources for the microservices can be easily accomplished as cloud native applications enjoy a great degree of resource scalability. The initial (and inefficient) resource allocation can be achieved using existing rule-based resource managers \cite{kubernetes_autoscaler}.}
Using this opportunistic resource reduction approach, \ouralg avoids causing intentional SLO violations as it always allocates enough resources for microservices, even when performing poorly (i.e., missing resource reduction opportunities).  \camera{To enable \ouralg's approach, we introduce the notion of ``monotonic resource reduction'' where we either reduce the resource of a microservice or keep it unchanged. In contrast, a non-monotonic resource reduction can be made through resource reduction for some microservices and resource increase for some other microservices with an overall total resource reduction (i.e., a greater total reduction than total increase).}
We observe that monotonic resource reductions result in a monotonic increase in response time.
Hence, we can use the response time as feedback to identify resource reduction opportunities to make gradual monotonic resource changes to reach efficient allocations. In addition, based on experiments on prototype microservice implementations, we identify that we can avoid resource reduction in bottleneck services using only two microservice-level performance metrics - CPU utilization and \camera{CPU throttling time}.

Our feedback-based design also allows us to seamlessly adapt a workload-aware design where we implement a novel approach of using dynamic workload ranges with a dynamic response time target. More specifically, to avoid time-consuming learning of the efficient allocation for different workload levels independently, we use dynamic ranging where \ouralg starts resource allocation for a large workload range (e.g., 100$\sim$1000 requests-per-second) and then gradually splits them into smaller ranges (e.g., 100$\sim$200 requests-per-second). We retain the resource allocation learned by the parent workload range during the range split to bootstrap the tuning for the new workload range. Based on the workload, we also dynamically alter the feedback from response time to allow headroom for response time change due to workload change.

Our performance evaluation reveals that \ouralg can attain resource efficiency close to the optimum \footnote{\camera{Optimum resource allocation refers to the minimum resource required to satisfy SLO. We describe how we identified the optimum resource allocation in Section~\ref{sec:benhmark}.}} with high probability. We also show that \ouralg can save as much as 33\% resource compared to rule-based resource allocation strategies of commercially available cluster managers. We demonstrate that \ouralg can seamlessly adapt to changes in microservice deployment due to changes in underlying cloud hardware. Moreover, we show that adaptability of \ouralg allows its integration with opportunistic resource management where variable SLO is used for trading performance for resource savings.

\noindent\textbf{Experimental methodology and artifact availability.}
We use three prototype microservices implementations widely used in academic research on microservices \cite{firm,gan2021sage,zhang2021sinan}. We implement \trainT from \cite{train_ticket} consisting of 41 microservices, \socS from \cite{sock_shop} with 13 microservices, and \hotelR from \cite{deathstarbench} with 18 microservices. We deploy these services in Docker \cite{docker} containers managed by Kubernetes \cite{kubernetes}. Our Kubernetes cluster consists of five nodes with one master node and four worker nodes. Each node is equipped with two 10-core Intel Xeon processors, and 128 GB of Memory running the Ubuntu 20.04.3 operating system. 
\camera{Our software artifacts are available at our GitHub repository \cite{pema}.}

\noindent\textbf{Limitations of the proposed approach.} 
We share our insight on the limitations of \ouralg on two different fronts - the fundamental limitations in \ouralg's design approach and the limitations of \ouralg's current implementation. Due to its non ML-heavy approaches, \ouralg's design loses on capturing complex interdependencies between microservices, and therefore is limited on the absolute best resource efficiency it can achieve. However, \ouralg makes up for this loss of optimization potential through its simplicity and adaptability to change (e.g., workload variation). Also, due to our randomized exploration based search, \ouralg offers provably efficient management and can result in arbitrarily inefficient resource allocations at times. 
We defer the discussion on the limitations of \ouralg's current implementation to the end of our paper in Section~\ref{sec:conclusion} to make it more meaningful to the reader.
\section{Preliminaries}

\subsection{Microservice Prototypes}

\begin{figure}[t]
\includegraphics[width=0.4\textwidth]{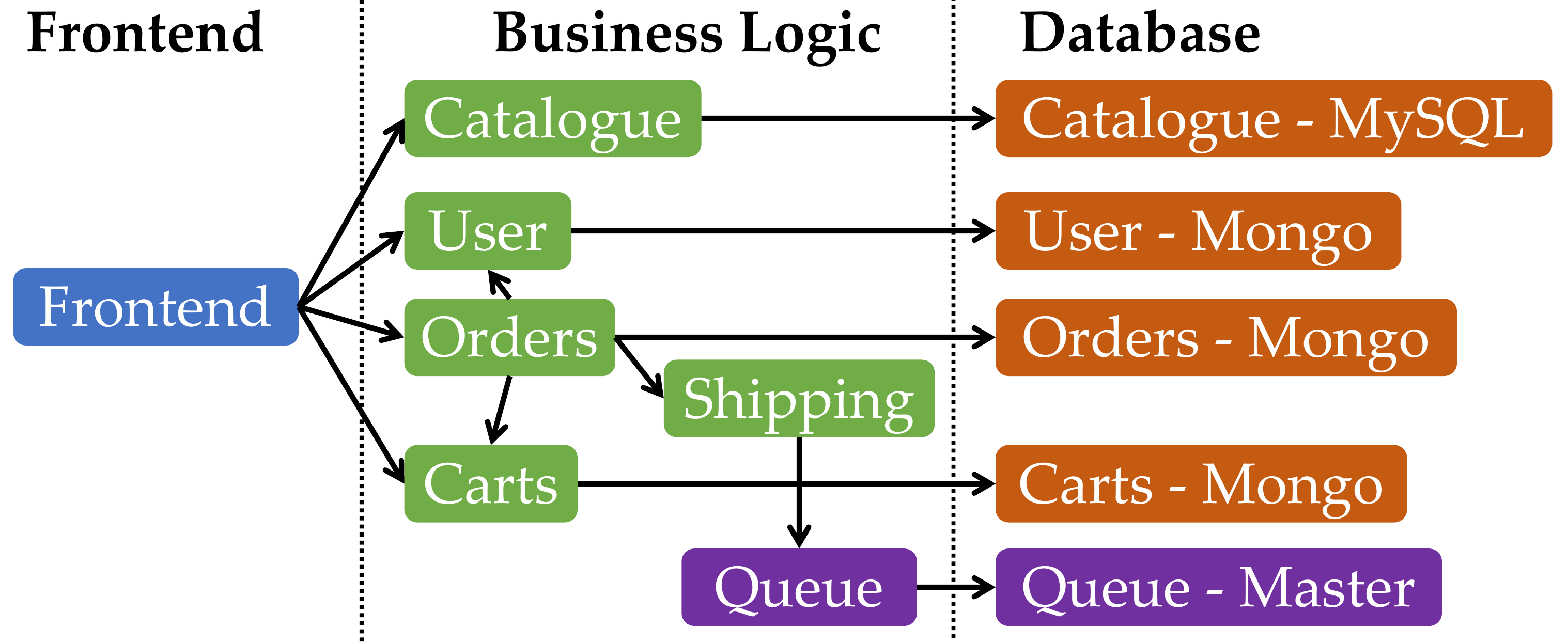}
\vspace{-0.3cm}
\caption{Architecture of the \socS \cite{sock_shop}.}
\label{fig:sock_shop_logic_diagram}
\end{figure}

\textbf{\socS \cite{sock_shop}.}
\socS implements the user-facing microservices of an e-commerce website. \socS's functionalities include searching, order placement, and shipping. Its functionalities can be divided into three parts - front-end, business-logic, and databases. The user requests arriving at the front-end are routed to appropriate microservices to serve the requests. The business-logic interact with each other and the databases as needed. 
The front-end is implemented using NodeJS, orders and carts microservices are implemented using Java, and the rest of the services are implemented with Go. Shipping service uses RabbitMQ to propagate messages to Queue-Master which is implemented in Java. The databases are implemented using MySQL and MongoDB. For \socS, we set the SLO response time to 250 milliseconds. 
The overall architecture is shown in Fig.\ref{fig:sock_shop_logic_diagram}. 

\begin{figure}[t]
\includegraphics[width=0.47\textwidth]{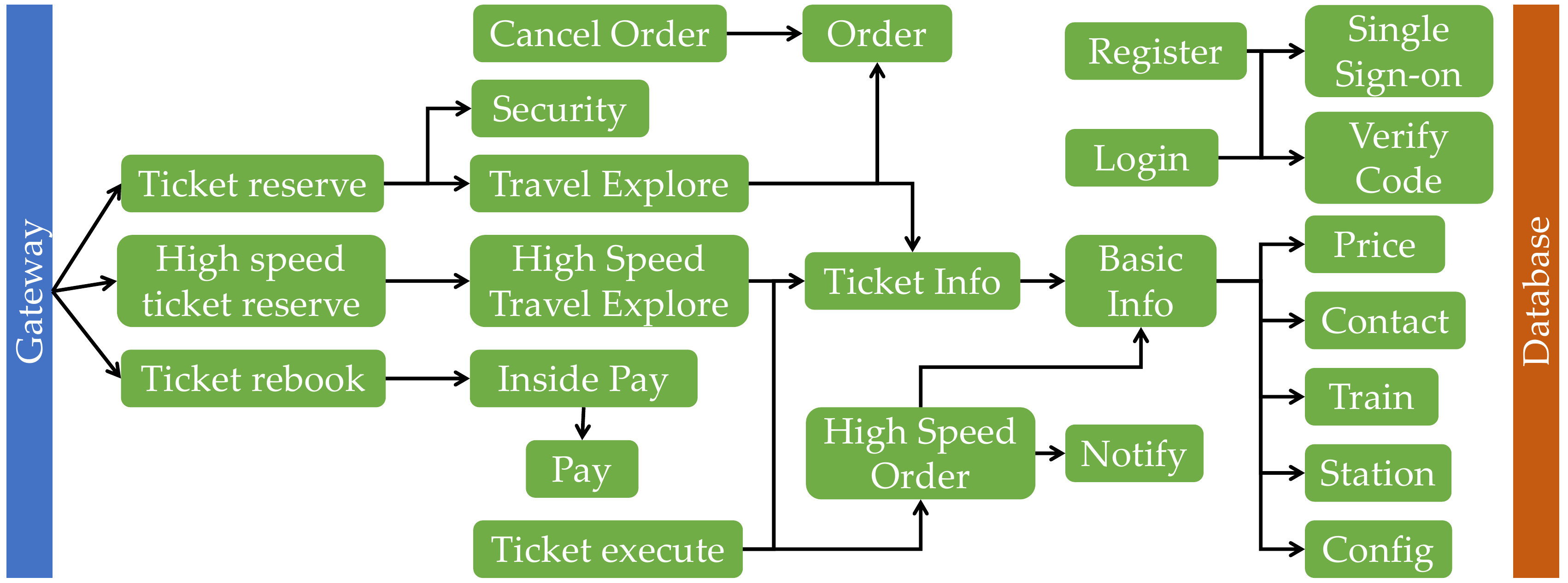}
\vspace{-0.3cm}
\caption{Architecture of the \trainT \cite{train_ticket, train_ticket_figure}.}
\label{fig:train_ticket_logic_diagram}
\end{figure}

\textbf{\trainT \cite{train_ticket}.}
\trainT implements a complete train ticket booking system consisting of 41 microservices. Its functionalities include ticket search with date and destination filtering, seat booking, ordering food, payment, and consignment service. The business logic of \trainT is implemented using 24 microservices divided into five layers where the microservices in the upper layers depend on the microservices of the lower layers. There are some intra-layer communications as well. The overall architecture is shown in Fig. \ref{fig:train_ticket_logic_diagram}. 
\trainT covers many features of microservices such as synchronous invocations, asynchronous invocations, and message queues. The \trainT business logics and front-end are built using NodeJS, Java, Python, and Go. The databases are implemented using MongoDB, and MySQL. For \trainT, we set the SLO response time to 900 milliseconds.

\textbf{\hotelR \cite{deathstarbench}.}
\hotelR application is adopted from DeathStarBench microservices benchmark applications. It has 18 microservices.
\hotelR lets users get nearby hotel information and reserve rooms. All the services in \hotelR are written in Go, and they communicate with each other via gRPC \cite{grpc}. The back-end uses Memcached for in-memory caching to provide faster searches while the persistent databases are implemented using MongoDB. 
The application is pre-populated with 80 hotels and 500 registered users. This application consists of 18 microservices. For \hotelR, we set the SLO response time to 50 milliseconds.  

\begin{figure}[t!]
\includegraphics[width=0.4\textwidth]{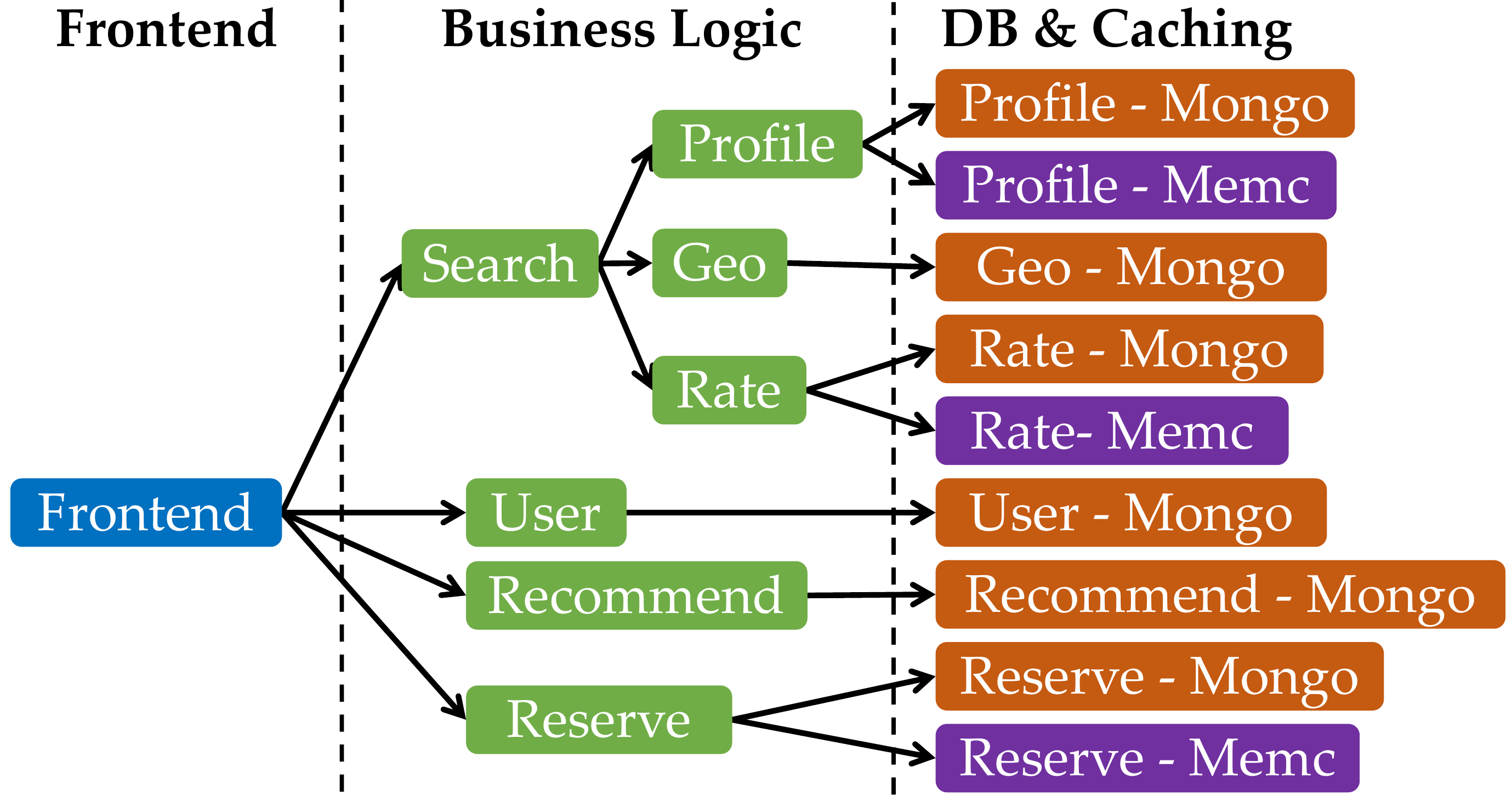}
\vspace{-0.3cm}
\caption{Architecture of the \hotelR \cite{deathstarbench}.}
\label{fig:hotel_reservation_logic_diagram}
\vspace{-0.3cm}
\end{figure}

\subsection{Performance Monitoring and Resource Allocation}

For performance monitoring of our container-based microservice implementation, we use Prometheus \cite{promethues} to collect container-specific metrics such as CPU utilization and CPU throttling. 
For collecting end-to-end latency performance and workload (i.e., requests per second), we use Linkerd \cite{linkerd}. 
We also use Jaeger \cite{jaeger} which provides detailed tracing of each request showing its service path through different microservices. Note that, our resource manager does not utilize Jaeger.

We use the 95-th percentile end-to-end response latency as a performance metric and refer to it as the application performance unless specified otherwise. \camera{For our cloud-based microservice applications which exploit request-level-parallelism, end-to-end response latency is the popular choice of performance metric \cite{jindal2019performance}.}
For microservice resources, we only consider the total CPU allocation to a microservice with the assumption that the memory is not a bottleneck resource. Furthermore, we do not explicitly address the number of container replicas and consider homogeneous settings for each microservice.

\subsection{Challenges in Microservice Resource Management}

\begin{figure}[t!]
\subfigure[\trainT]{\includegraphics[width=0.15\textwidth]{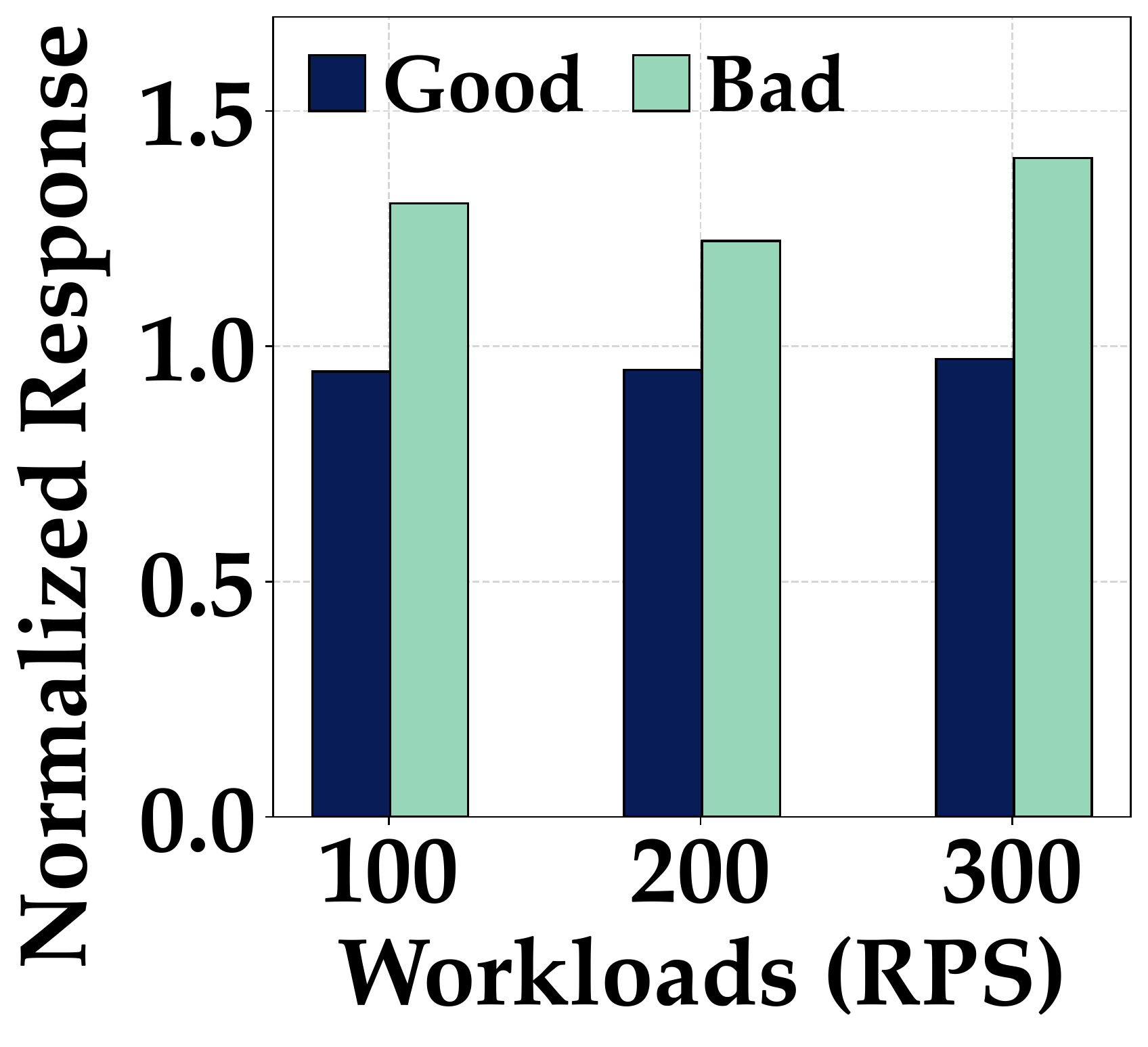}}\hspace{0.1cm}
\subfigure[\socS]{\includegraphics[width=0.15\textwidth]{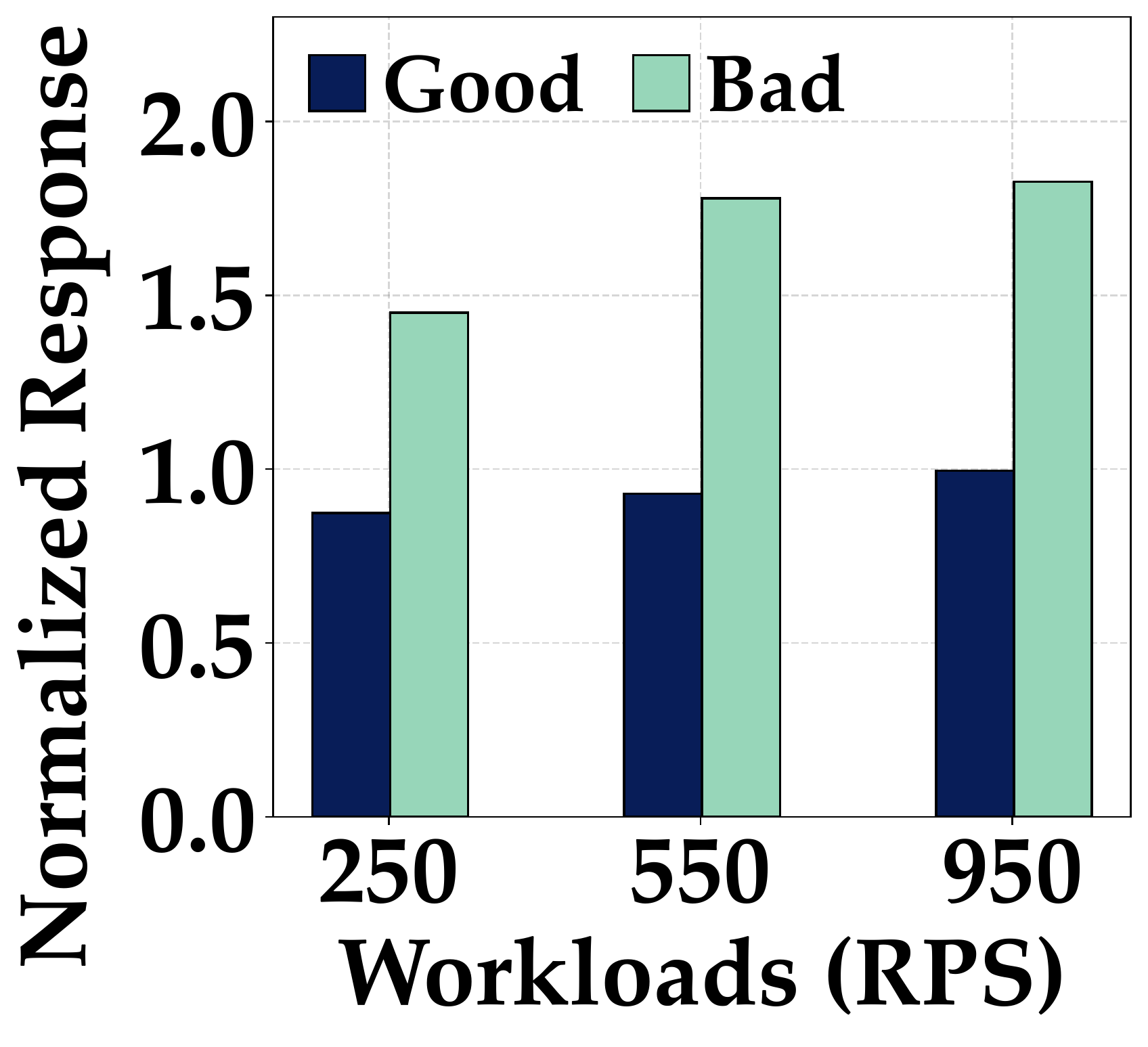}}\hspace{0.1cm}
\subfigure[\hotelR]{\includegraphics[width=0.15\textwidth]{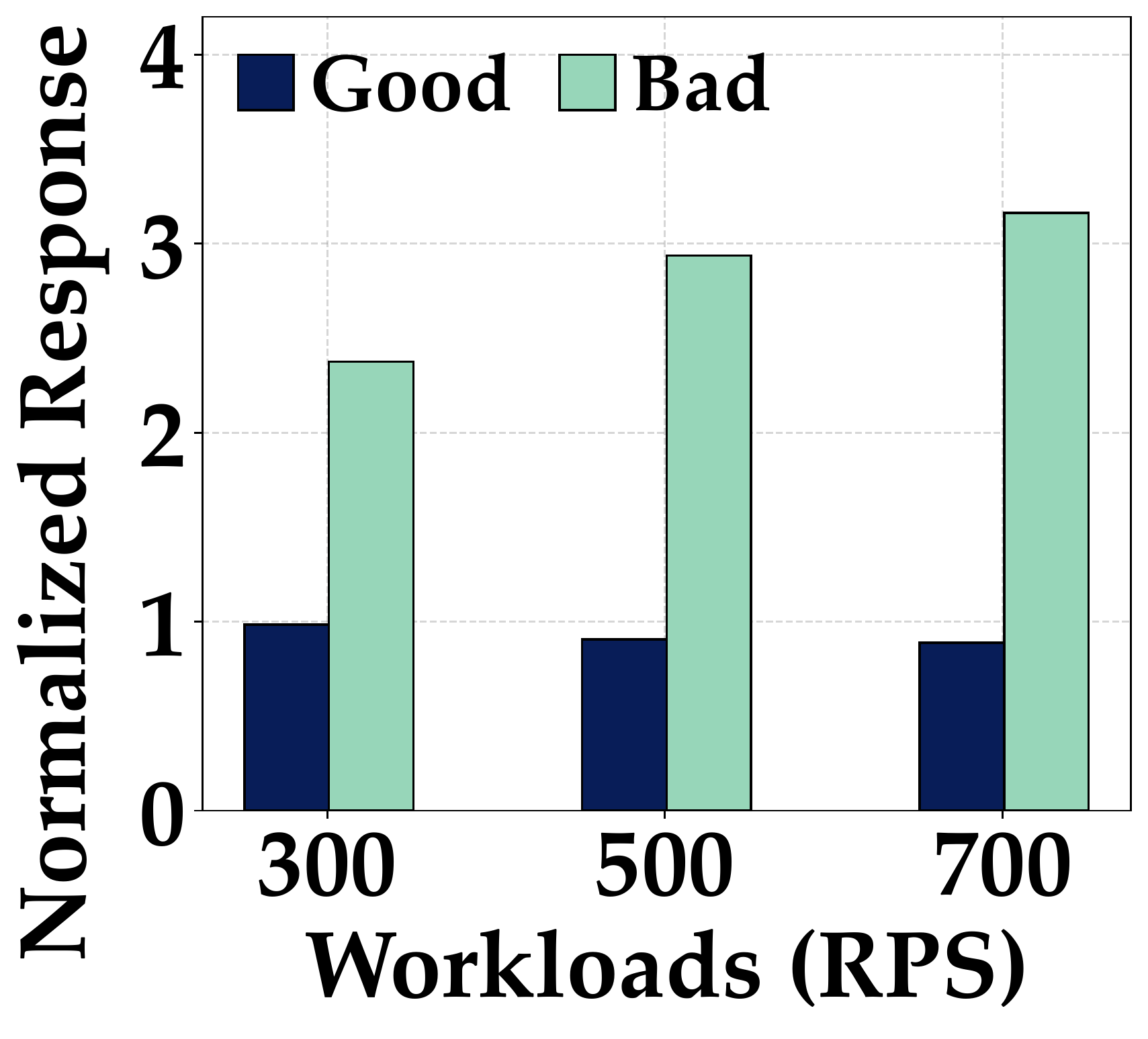}}
\caption{Impact of ``Good'' (i.e., satisfies SLO) and ``Bad'' (i.e., violates SLO) resource distribution on the response time normalized to the SLO at different workloads levels. 
In Fig. (a), for workloads 100, 200, and 300, total CPU allocations are 40.5, 42, and 47 respectively. In Fig. (b), for workloads 250, 550, and 950, total CPU allocations are 6.3, 7.7, and 14.1, respectively. In Fig. (c), for workloads 300, 500, and 700, total CPU allocations are 5.1, 6.9, and 9.4, respectively.
}
\label{fig:good_vs_bad_configurations}
\vspace{-0.5cm}
\end{figure}

As in any general computing system, the performance of microservices applications depends on their resource allocation. Various theoretical and practical tools have been developed over the years to establish a mathematical relationship between computing resource and performance \cite{gandhi2012autoscale}. However, they are not equipped to capture complex interactions between different microservices. Any request's end-to-end response time (i.e., performance) is the aggregation, often non-linearly due to parallel processing, of time spent in many microservices. Consequently, the presence of any microservice with a resource bottleneck on the service path affects the end-to-end response time. Meanwhile, the resource demand for different microservices can be widely different based on their service. Hence, the distribution of resources among different microservices plays a crucial role in application performance.

To demonstrate the importance of resource distribution, we run a few experiments on our microservices prototypes. We first identify ``good'' resource allocations that satisfy the SLOs for the prototypes for different workload levels. We then change these to ``bad'' distributions by randomly altering resource allocations while keeping the total resource the same. Fig.~\ref{fig:good_vs_bad_configurations} shows the impact of this resource distribution -  even with the same amount of resources, the performance varies significantly because of changes in distribution. \camera{For, \trainT we see as much as $43.88\%$ increase in response time while \socS and \hotelR suffer up to $91.3\%$ and $256.2\%$ increase, respectively.}

\begin{figure}[t!]
\subfigure[CPU allocation]{\label{fig:ss_good_vs_bad_configurations}\includegraphics[width=0.23\textwidth]{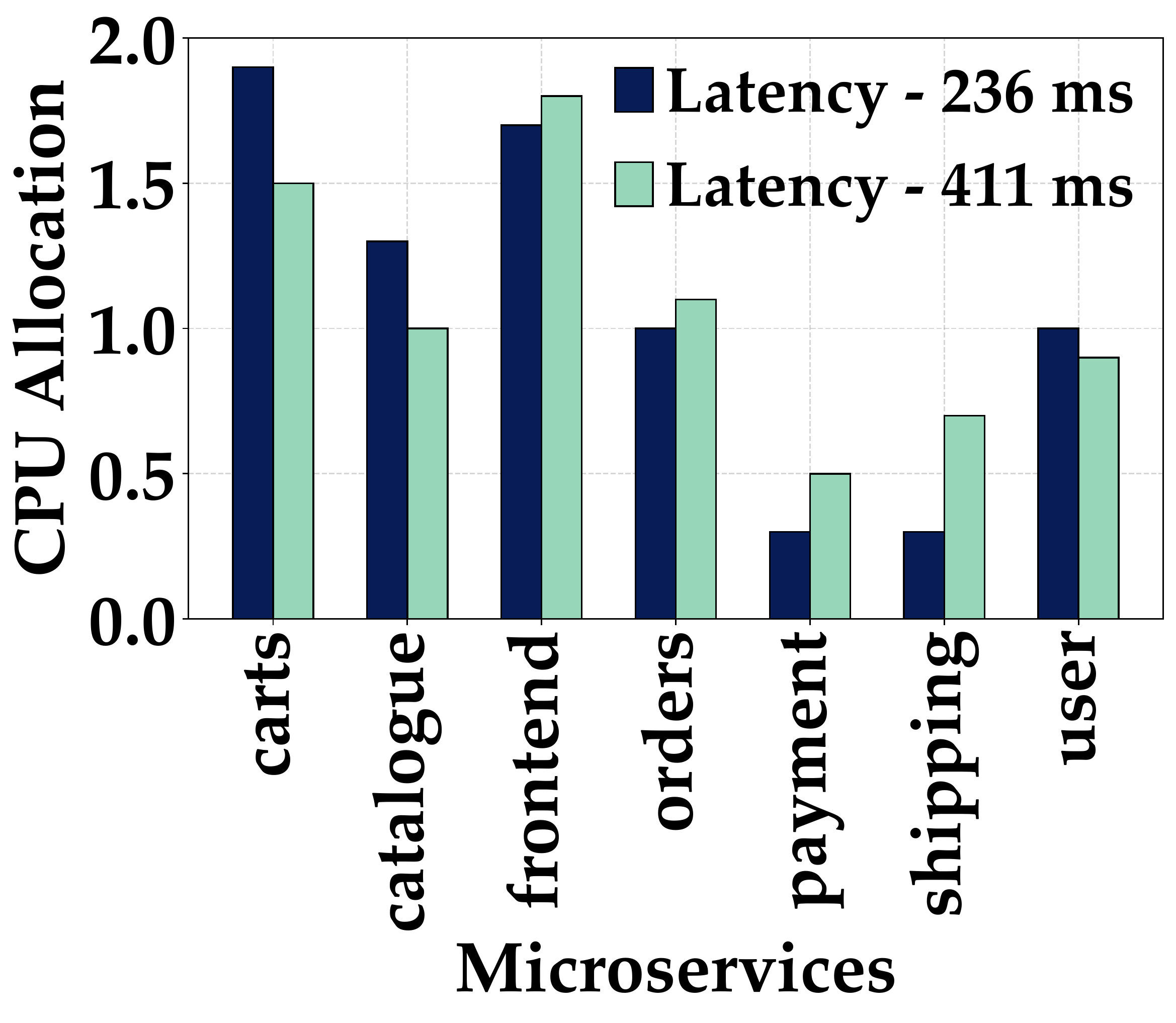}}\hspace{0.2cm}
\subfigure[Utilization]{\label{fig:ss_good_vs_bad_utilizations}\includegraphics[width=0.23\textwidth]{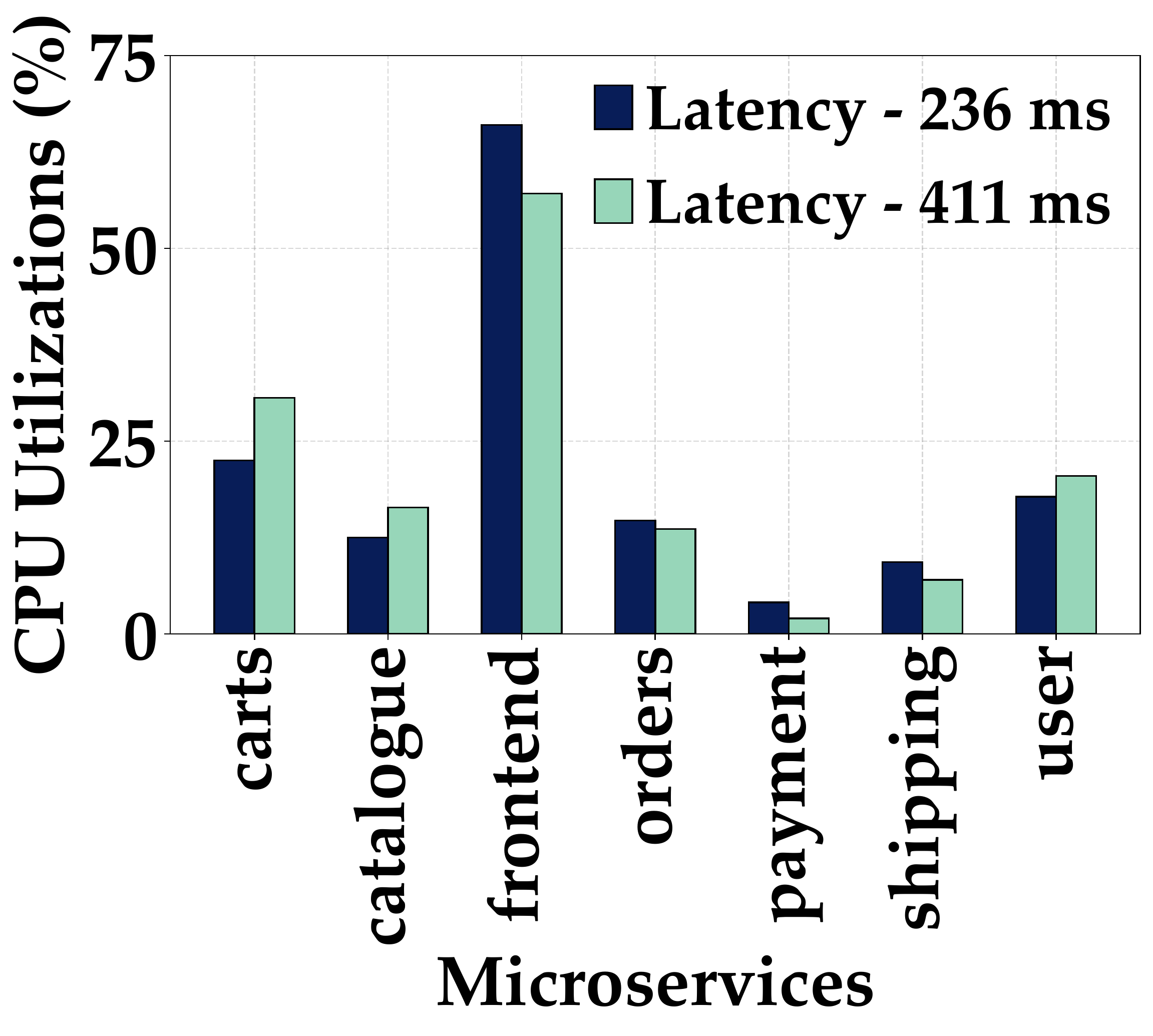}}
\caption{(a) Total CPU allocation of 7.5 distributed among different microservices of \socS. (b) CPU utilization.}
\end{figure}

Due to the large configuration space, the ``good'' resource distribution cannot be readily determined for microservices. Also, the nature of processing done in different microservices is different and cannot adhere to any general resource allocation principle, such as keeping utilization lower than a certain level \cite{azure_autoscale, aws_autoscale, google_autoscale, kubernetes_autoscaler}. To illustrate this, we show the resource distributions of \socS's microservices for the good and bad configurations with the same amount of total resource in Fig.~\ref{fig:ss_good_vs_bad_configurations} and the corresponding CPU utilization in Fig.~\ref{fig:ss_good_vs_bad_utilizations}. We see that there is no readily identifiable root cause (e.g., microservice with bottleneck resource) in response latency in Fig.~\ref{fig:ss_good_vs_bad_utilizations} for the 74\% increase (236 milliseconds to 411 milliseconds). Also, while we see an increase in utilization for the \texttt{cart}\xspace, \texttt{catalogue}\xspace, and \texttt{user}\xspace services for the bad configuration, their utilization remains below the \texttt{frontend's}\xspace utilization, making it impossible to employ any common utilization-based resource allocation policy.
Furthermore, we see that the utilization change due to resource change is different for different services. For example, the \texttt{frontend's}\xspace utilization changed more than \texttt{orders}\xspace even though they experienced similar resource change. This indicates that resource allocation policies that try to increase overall utilization \cite{firm}, may not be the most efficient.

\camera{To summarize, \emph{for efficient microservice management, it is crucial to identify how resources should be distributed among different microservices as the same amount of resources can result in significantly different performance based on which microservice gets how much resources. However, finding the efficient resource distribution is very hard as there are no easily generalizable markers (e.g., high utilization) to assist in the resource allocation.}}

\section{Design of \ouralg}

We have two design goals for our resource manager - (1) assure QoS (i.e., avoid SLO violations), and (2) find efficient resource allocation. 
Using a discrete-time model with a time step $\Delta t$ (e.g., one minute) where the microservice resource allocation decisions are updated at the beginning of each time step, we formalize our resource management as the following
optimization problem 
\pone (\textbf{O}ptimum \textbf{R}esource \textbf{A}llocation)
\begin{align}
\mathsf{ORA}:  \underset{\textbf{x}^t} {\text{minimize}}  \quad & \sum_{i=1}^{N} x_i^t \label{equation:opt}\\
 \text{subject to} \quad& \mathcal{F}(\textbf{x}^t) \leq R
\end{align}
Here, at time step $t$, $\textbf{x}^t = (x_1^t,x_2^t,\cdots,x_N^t)$ is the resource allocation vector of the $N$ microservices, $\mathcal{F}(\textbf{x}^t)$ is the end-to-end latency response of the application for resource allocation $\textbf{x}^t$, and $R$ is the response latency threshold defined in the SLO. 
In what follows, we develop \ouralg (\textbf{P}ractical \textbf{E}fficient \textbf{M}icroservice \textbf{A}utoscaling) - a practical microservices resource manager that finds a provably efficient solution to \pone.
We first discuss the design principles of \ouralg to achieve our goals (i.e., the solution to \pone), followed by the rationale for our choices and implementation details of \ouralg.

\camera{Note here that, instead of minimizing the total resource allocation, \pone can also adopt cost minimization as its goal by replacing $x_i^t$ in Eqn.~\eqref{equation:opt} with $\mathcal{C}(x_i^t)$ which represents the cost of resource $x_i^t$. Moreover, resource allocation vector $\textbf{x}^t$ is not restricted to CPU allocations only. We can incorporate other types of cloud resources such as memory allocation and I/O bandwidth in $\textbf{x}^t$. Nevertheless, our general solution principle still applies, albeit the opportunistic resource reductions need to be conducted on multiple resource dimensions.}

\subsection{Design Principles of \ouralg}

\textbf{A learning-based approach.} Achieving either of our design goals for a microservice-based application is non-trivial due to their complex topology and inter-dependency between different microservices. Moreover, the relation and interaction with each other for these microservices varies with applications and deployments, even among different versions of the same application. Not to mention, the underlying cloud hardware (e.g., processor type/model) hosting these applications also affects the microservice performance and resource allocation. Consequently, our resource manager needs to identify resource allocation strategies for each microservice implementation and at the same time be able to adapt as the application evolves. 
Hence, we take a learning-based approach where \ouralg iteratively interacts with the application through a feedback loop to navigate towards efficient resource allocations.

\textbf{Provably efficient resource allocation.} 
Solving \ouralg can be interpreted as tuning the application resources that will make the response latency exactly equal to the SLO specified level.
However, since the resource distribution across different microservices affects the latency and microservice-based applications usually consist of many microservices, 
there could be many different resource allocations that result in a latency equal to the SLO.
Consequently, in \ouralg, instead of finding the best resource allocation (i.e., the lowest aggregate resource), our goal is to find a resource allocation close to the optimum with fewer iterations.

\textbf{QoS preserving learning.} An unwanted pitfall of the learning-based approach in the existing literature is that the system needs to learn ``bad'' resource allocations that cause SLO violation by causing/creating these violations \cite{firm,zhang2021sinan,gan2019seer,gan2021sage}. While our approach too cannot completely eradicate the possibility of SLO violations, unlike prior works, we do not cause them intentionally. 
\camera{Instead, we adopt a QoS conservative approach where we start from with sufficient resource for all microservices to satisfy SLO, and then iteratively search for resource reduction opportunities based on the application's performance statistics.}
During the search/learning, \ouralg always tries to maintain latency performance better than the SLO.  
Moreover, we dynamically tune how much resource we reduce based on how close our performance is to the SLO and stop tuning if the performance is at the SLO level. For example, with a response time SLO of 250ms, \ouralg will try to reduce more resources when the response time is 150ms than when the response time is 200ms. Hence, during resource allocation navigation, \ouralg does not set a resource allocation to violate the SLO intentionally.

\textbf{Feedback-based navigation.} Starting with ample resources for each microservices to comfortably satisfy SLO, \ouralg uses the difference between current application performance and the SLO as an indicator of resource reduction opportunity. However, it does not tell us on which microservice(s) we should exercise the resource reduction. Hence, \ouralg uses microservice-wise performance metrics to determine the target microservices. More specifically, \ouralg uses the microservice-wise performance metrics to filter out the microservices approaching their bottleneck resource configuration and then implements a randomized selection process where the probability of picking a microservice is determined by its performance metrics. With unknown relation between a microservices resource allocation with the overall application performance, a guided randomized selection allows \ouralg to explore various possible combinations of resource allocation.

\subsection{Supporting Results for Design Rationales}

Here, we provide corroborating observations for \ouralg's design using our prototype microservices implementations. We first show why application's performance can be a safe yet effective indicator of resource reduction, followed by how microservice-wise performance metrics can help \ouralg navigate.

\begin{figure}[t!]
\subfigure[CDF of latency change]{\label{fig:monotonic_cpu_changes_cdf}\includegraphics[width=0.23\textwidth]{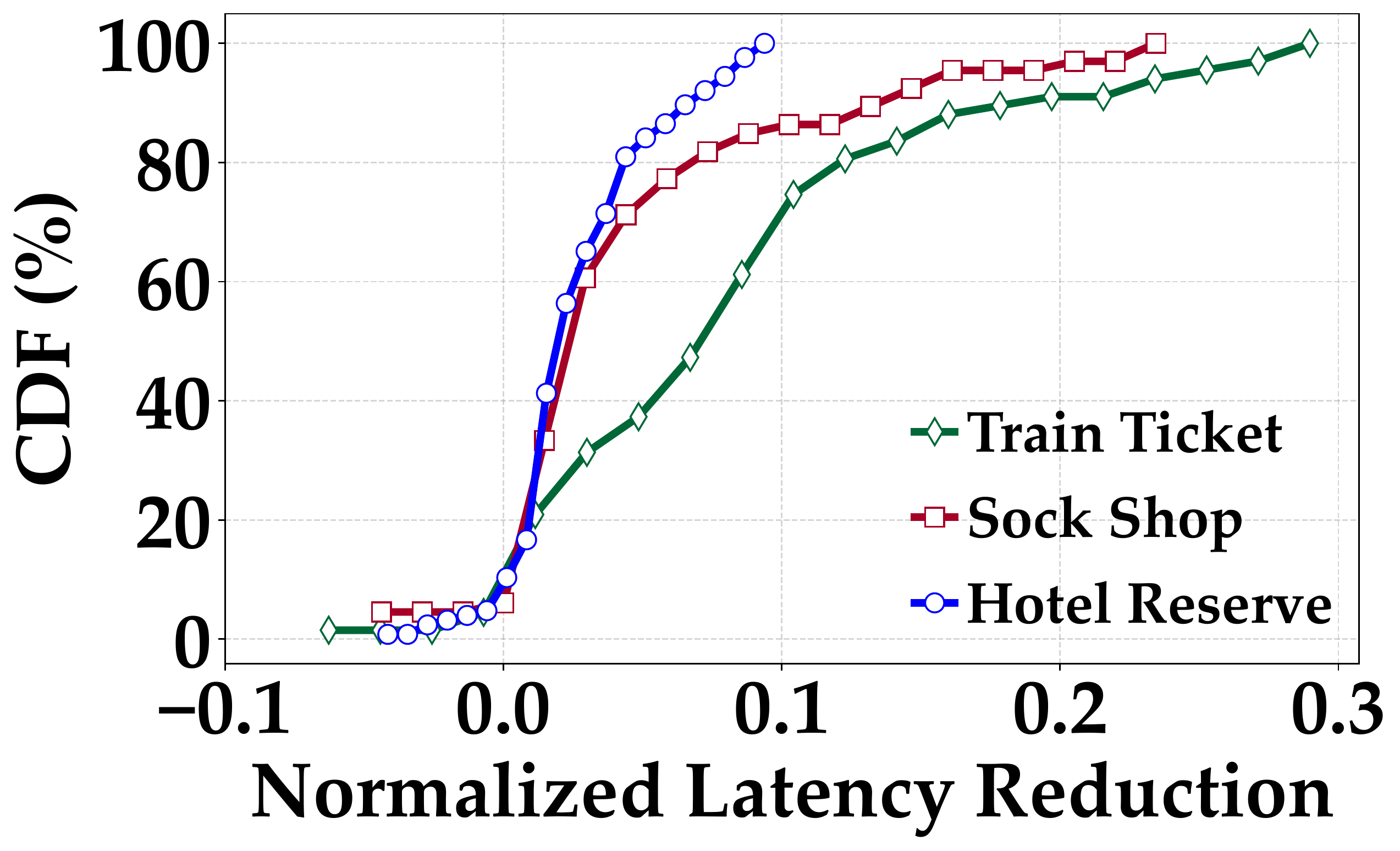}}\hspace{0.2cm}
\subfigure[Response time change]{\label{fig:delta_vs_cost_3}\includegraphics[width=0.23\textwidth]{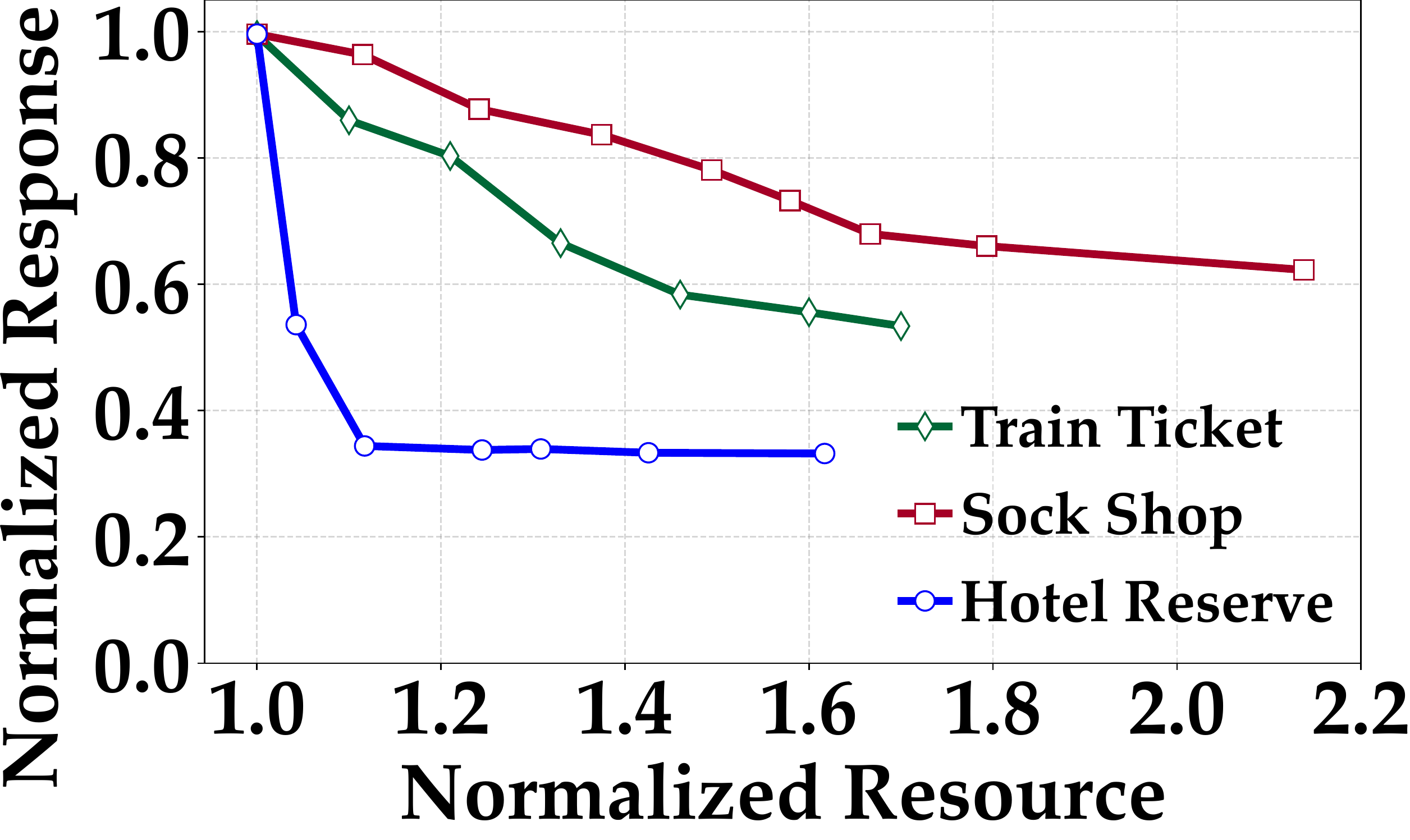}}
\caption{(a) Distribution of \camera{end-to-end response time increase} (normalized to SLO) due to monotonic resource reduction. (b) Change in response time (normalized to SLO) with resource (normalized to optimum).}
\end{figure}

\textbf{Gradual resource reduction for efficiency.}
In \ouralg, we use the difference between SLO specified response time and current system response time to determine how much resource-saving opportunity is available.
Our design choice is motivated by our observation that, in general, \textit{monotonic resource changes across microservices result in monotonic changes in the end-to-end response time}. We say a resource reduction is monotonic if some microservice resources are decreased while other microservices' resources are unchanged. A resource change is not monotonic if some microservices receive greater resources while some others have their resource reduced, regardless of what happens to the aggregate resource allocation. Fig.~\ref{fig:monotonic_cpu_changes_cdf} shows the CDF of \camera{increase in end-to-end response time for monotonic resource reduction for our applications. Note that there is no direct relationship between resource reduction and the amount of change in response time. This is because, the same amount of resource reduction on different microservices will have different impact on the end-to-end response time. The CDF is showing distribution of latency increase for random amounts of monotonic resource reduction on random numbers of microservices at random initial (before resource reduction) resource allocations. The CDF highlights the most likely impact of a monotonic resource reduction - an increase in the response latency regardless of the state of the microservice, i.e., its total resource allocation.}
\camera{The CDFs also show that the opposite, i.e., response latency decreasing with resource reduction, happens an only handful of times ($10.2\%$ for \trainT and $6.1\%$ for \socS).} 
We attribute these cases as transient anomalies based on our observation of the application's performance metric fluctuations.

\camera{The key take away from Fig.~\ref{fig:monotonic_cpu_changes_cdf} is that \emph{by making monotonic resource reductions, we can gradually increase the latency to the SLO level.}} In Fig.~\ref{fig:delta_vs_cost_3}, we show examples of such monotonic resource reduction steps and its impact on latency. Here, we normalize the resource to the optimum resource allocation and the latency to the SLO level. \ouralg's goal in Fig.~\ref{fig:delta_vs_cost_3} is to reach coordinate (1, 1) by gradually making monotonic resource changes. \camera{Note that the resource reduction steps in Fig.~\ref{fig:delta_vs_cost_3} is not unique. Moreover, monotonic resource reduction alone does not guarantee to reach the optimum resource allocation keeping the response latency within the SLO. Instead, it offers a QoS preserving approach of navigation to find efficient resource allocation.}

\textbf{Microservice-wise augmentation.}
While the response latency tells us about the resource reduction opportunities, it does not tell us from which microservices we should reduce the resources. We need to avoid microservices that may create a bottleneck during this resource reduction. \camera{We define a microservice's ``bottleneck resource'' as the resource allocation that makes the microservice a bottleneck.} In \ouralg, we use microservice-level performance metrics to identify the microservices with imminent bottleneck resources. However, as opposed to prior works where complex machine learning models are applied to determine such bottleneck services, we use only two performance metrics -  CPU utilization and CPU throttling time \cite{cpu_throttle}. 

Our choice of these performance metrics is based on our experiments. We intentionally create bottlenecks and use feature extraction to identify which performance metrics can be used to identify the bottleneck services reliably. Note that these experiments are done to assist in our design. \ouralg does not need any offline experiments or pre-training. For each microservice, we collect the following performance metrics - \texttt{cpu\_usage\_seconds\_total}, \texttt{memory\_usage\_bytes}, \texttt{cpu\_cfs\_throttled\_seconds\_total}, Jaeger tracing - \texttt{self\_time}, and \texttt{duration}. We then run classification with various combinations of the performance metrics as features. We find that, when used as the classification features, CPU utilization and CPU throttling time give us the highest classification accuracy. Table \ref{table:performance_metrics} shows the classification accuracy for different applications with various bottleneck services.

\begin{table}[t!]
\begin{center}
\caption{Classification accuracy with CPU utilization and CPU throttling time as features to detect bottleneck microservices.}
\resizebox{1\linewidth}{!}{%
\begin{tabular}{|c|c|c|}
\hline
\textbf{Microservice Name} & \textbf{Botteleneck Services} & \textbf{Accuracy (\%)} \\ \hline
\trainT & \texttt{seat} & 94.18    \\ \hline
\trainT & \makecell{\texttt{seat, ticketinfo}} & 96.2 \\ \hline
\socS & \texttt{carts} & 100.0 \\ \hline
\socS  & \makecell{\texttt{carts, orders}} & 98.3 \\ \hline
\hotelR & \texttt{front-end} & 97.8 \\ \hline
\hotelR & \makecell{\texttt{front-end, search}} & 95.6 \\ \hline
\end{tabular}}
\label{table:performance_metrics}
\end{center}
\end{table}

\begin{figure}
\subfigure[CPU utilization]{\label{fig:ts_bottleneck_utils}\includegraphics[width=0.23\textwidth]{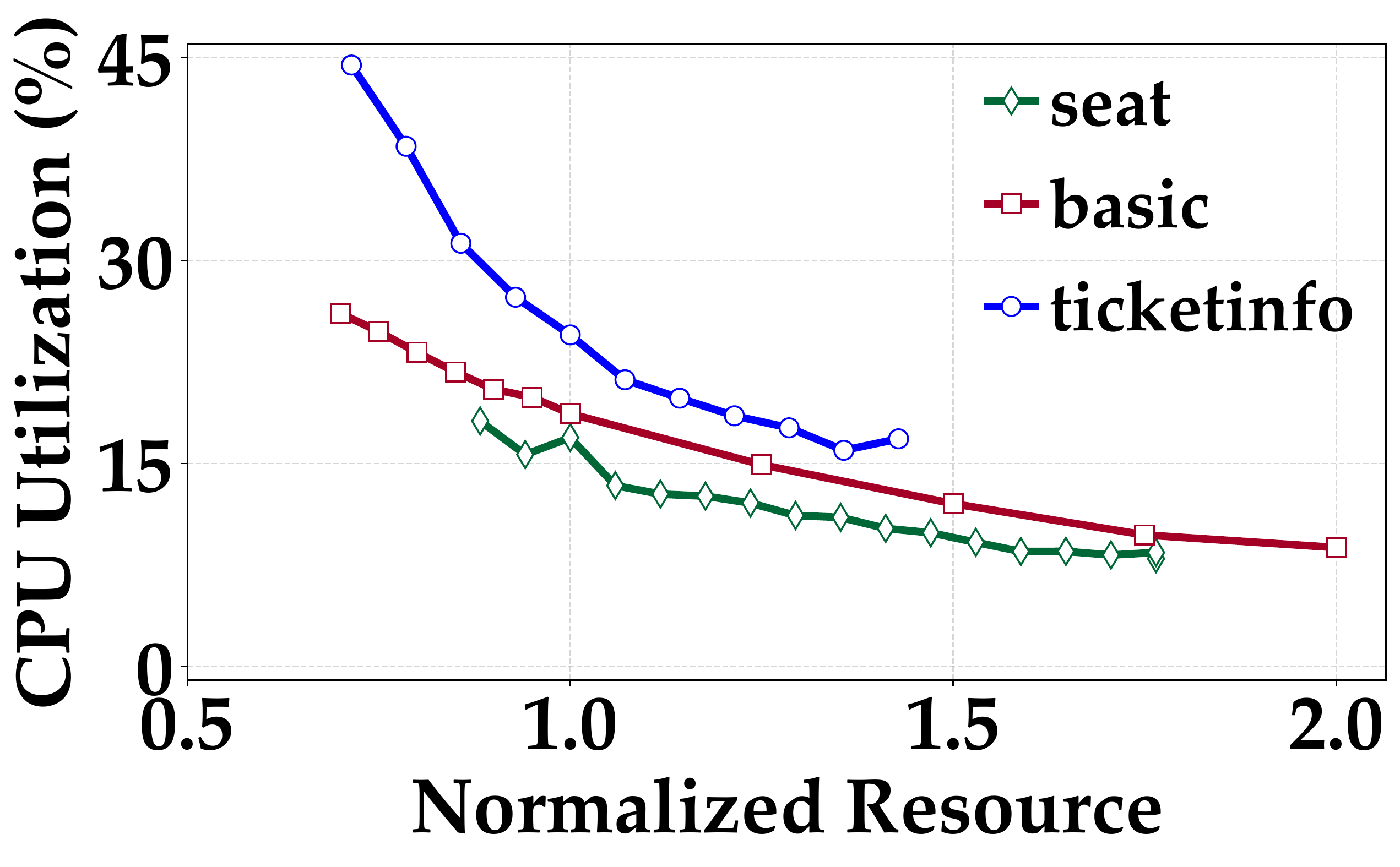}}\hspace{0.2cm}
\subfigure[CPU throttling time]{\label{fig:ts_bottleneck_throttles}\includegraphics[width=0.23\textwidth]{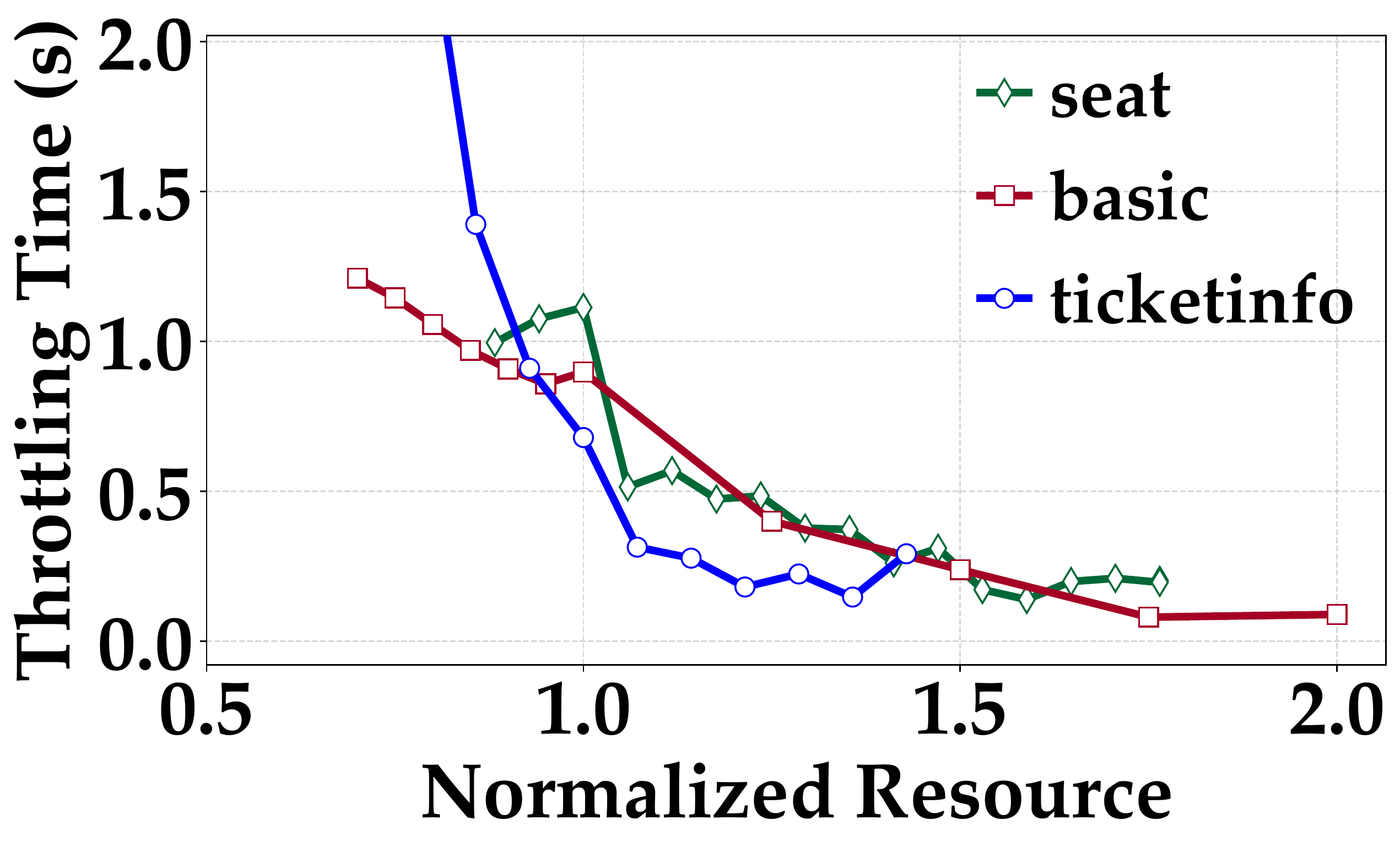}}
\caption{Changes in CPU utilization and CPU throttling time with resource allocation for three bottleneck microsservices in \trainT -  \texttt{seat}, \texttt{basic}, and \texttt{ticketinfo}. 
}
\label{fig:ts_bottlenecks}
\end{figure}

\camera{To better understand the role of CPU utilization and CPU throttling time as bottleneck indicators, we track these metrics for three different microservices in \trainT - \texttt{seat}, \texttt{basic}, and \texttt{ticketinfo}, as we reduce their resources to create bottlenecks.} To identify the bottleneck, we allocate sufficient resources to all other microservices. Fig.~\ref{fig:ts_bottlenecks} shows the change in CPU utilization and CPU throttling as we reduce the resource of the microservice under investigation. We normalize the microservice resource allocations to their respective bottleneck resources. We make a few important observations here. \textit{First,} the CPU utilization (Fig.~\ref{fig:ts_bottleneck_utils}) changes gradually as the microservice approaches and eventually crosses the bottleneck resource. We also see that the utilization corresponding to bottleneck is different for different microservices. For example, ticketinfo's bottleneck utilization is around 25\%, whereas seat's bottleneck utilization is around 15\%. \textit{Second,} CPU throttling time changes rapidly at bottleneck resource. The bottleneck CPU throttling time also varies with microservices.

\subsection{\ouralg}

Here we present the details of \ouralg's implementation that builds on our design principles and experimental observations.

\textbf{Resource reduction opportunity.} 
In \ouralg, similar to gradient descent, we start with sufficient resources for all microservices and gradually decrease their resource based on how our resource change affects the end-to-end response time. 
We update resource allocation in regular intervals based on the response time observed in the previous interval. Since we rely on the response time statistics, we set sufficiently long update intervals to have stable response time statistics. For instance, in \trainT, \socS, and \hotelR, we use update interval of two minutes. 
For resource reduction at time step $t$, we first decide the number of microservices $n^t$ to reduce resources from using
\begin{align}\label{eq:ns_equation}
    n^t = N\cdot \min \left( \frac{R-r^{t-1}}{\alpha R},1\right),
\end{align}
where $r^{t-1} = \mathcal{F}(\textbf{x}^{t-1})$ is the response time in the previous time step. $\alpha\leq 1$ is a user-defined non-negative parameter that determines how aggressively we want to reduce the resource. 
A smaller $\alpha$ will reduce resource more aggressively and vice versa.

Next, using similar approach as Eqn.\eqref{eq:ns_equation}, we decide how much resource we reduce in the $n^t$ microservices in percentage using
\begin{align}\label{eq:ds_equation}
    \Delta^t = \beta \cdot \min \left( \frac{R-r^{t-1}}{\alpha R},1\right) \cdot 100\%,
\end{align}
where $\beta \leq 1$ is another user defined parameter that decides the maximum resource reduction for any microservice in one time step. A high value of $\beta$ makes \ouralg aggressively change the resource between update intervals and vice versa. We analyze the impact of $\alpha$ and $\beta$ in our evaluation in Section~\ref{sec:parametr_sensitivity}.

Using Eqns.~\eqref{eq:ns_equation} and \eqref{eq:ds_equation}, \ouralg dynamically adjusts the amount of monotonic resource reduction as our response time $r^t$ approaches SLO limit $R$. We can also set the values of $\alpha$ and $\beta$ dynamically to have more aggressive reduction when $R-r^{t-1}$ is high and reduce the amount of reduction per interval as $r^t$ approaches $R$. In addition, to avoid triggering resource change for transient perturbation in response time, we can keep a response time buffer by scaling down $R$, for instance, to 95\%, in Eqns.~\eqref{eq:ns_equation} and \eqref{eq:ds_equation}. 

\textbf{Avoiding bottleneck services.}
For the $i$-th microservice, we denote its utilization as $u_i$ with a bottleneck threshold $U_i^{th}$ and CPU throttling time as $h_i$ with a bottleneck threshold $H_i^{th}$.  To decide the $n^t$ candidate microservices, we first take the set of microservices that has a CPU throttling time less than their respective thresholds. We denote the set of indexes of these microservices as $\mathcal{I}^t = \{i: h_i^{t-1}\leq H_i^{th}\}$.
We then normalize the utilization of each microservice in $\mathcal{I}^t$ to their respective utilization threshold as $u^{* t-1}_i = \frac{u_i^{t-1}}{U_i^{th}}$ and  update the probability of each  microservice in $\mathcal{I}^t$ as follows
\begin{align}\label{eq:probability}
    p_i^t = 1-\frac{u^{*t-1}_i-\min_{i\in \mathcal{I}^t}(u^{*t-1}_i)}{1-\min_{i\in \mathcal{I}^t}(u^{*t-1}_i)}
\end{align}
Here, $\min_{i\in \mathcal{I}^t}(u^{*t-1}_i)$ means the minimum normalized utilization among all the microservices in $\mathcal{I}^t$. Eqn.\eqref{eq:probability} indicates that a microservice with utilization equal to its threshold, i.e., $u^{*t-1}_i=1$ will result in a ``zero'' probability ($p_i^t = 0$), whereas the microservice with the lowest utilization, i.e., $u^{*t-1}_i = \min_{i\in \mathcal{I}^t}(u^{*t-1}_i)$, will have the probability of ``one'' ($p_i^t = 1$). We populate a new candidate set $\mathcal{I}^{*t}$  with a inclusion probability of $p_i^t$ for the $i$-th microservice. If the size of  $\mathcal{I}^{*t}$ is equal to or smaller than $n^t$, we take the entire set $\mathcal{I}^{*t}$ and reduce each microservice in $\mathcal{I}^{*t}$ and reduce their resource by $\Delta^t$. However, if the size of $\mathcal{I}^{*t}$ is greater than $n^t$ we uniformly randomly choose $n^t$ microservices from  $\mathcal{I}^{*t}$.

\begin{figure}[t!]
\includegraphics[width=0.47\textwidth]{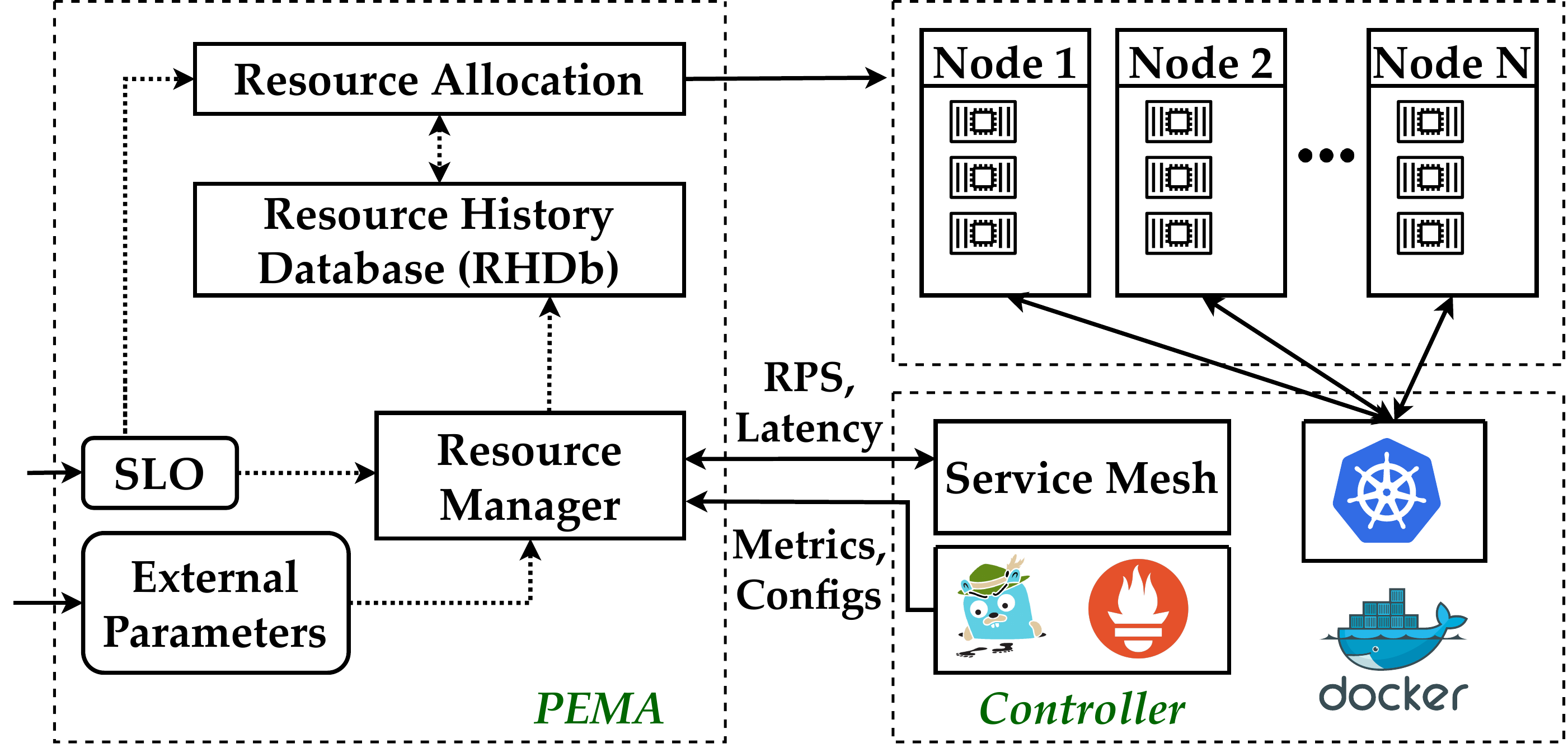}
\caption{Block diagram of the \ouralg.}
\label{fig:pema_system_design}
\end{figure}

\textbf{Dynamically updating bottleneck thresholds.} As shown in Fig.~\ref{fig:ts_bottlenecks}, the bottleneck thresholds for utilization and CPU throttling time varies among microservices. Hence, we need to learn the appropriate threshold settings for each microservice. In \ouralg, we begin with a conservative estimation of utilization threshold set at 15\% and CPU throttling time threshold of ``zero'' (i.e., no CPU throttling) for all microservices. We expect all microservices to satisfy these thresholds as \ouralg starts with ample resource allocation. Similar to our resource reduction approach, we opportunistically increase these thresholds. More specifically, at the beginning of every time step $t$, we update the utilization and CPU throttling time thresholds as follows
\begin{align}
    U_i^{th} = \max \left(U_i^{th}, u_i^{t-1} \right), \forall i \label{eq:util_th_update}\\
    H_i^{th}=\max\left(H_i^{th}, h_i^{t-1}\right),  \forall i \label{eq:cpu_th_update}
\end{align}

 \begin{algorithm}[t!]
 \caption{\ouralg} 
 \label{alg:main_algorithm}
 \begin{algorithmic}[1]
 \renewcommand{\algorithmicrequire}{\textbf{Input:}}
 \renewcommand{\algorithmicensure}{\textbf{Output:}}
    \REQUIRE SLO ($R$), affinity for resource reduction ($\alpha$), maximum resource reduction limit ($\beta$), bottleneck utilization ($U_i^{th}$), and bottleneck CPU throttling time ($H_i^{th}$) for all microservices, exploration probability parameters $A$ and $B$
    \ENSURE  Resource allocation ($\textbf{x}$) \\
    \FOR {each time-step $t$}
        \STATE \textbf{Performance metrics:} Collect end-to-end response time ($r^{t-1}$), CPU utilization $u_i^{t-1}$, and CPU throttling time $h_i^{t-1}$.
        
        \STATE \textbf{Database update.} Insert $x_i^{t-1}$, $r^{t-1}$, $U_i^{th}$, and $H_i^{th}$ to resource allocation history data base with key $t-1$.
        
        \STATE \textbf{Handling SLO violation.} If $r^{t-1} > R$, update resource allocation to configuration from the resource allocation database with minimum resource and no SLO violation. Go to Line~\ref{line:endFor}.
        
        \STATE \textbf{Updating bottleneck thresholds.} For all microservices, update bottleneck thresholds for utilization,  $U_i^{th}$, and CPU throttling time, $H_i^{th}$, following Eqns.~\eqref{eq:util_th_update} and \eqref{eq:cpu_th_update}, respectively.
        
        \STATE \textbf{Exploration.} With a probability $p_e^t$ defined in Eqn.~\eqref{eq:exploration}, update resource allocation, $\textbf{x}^t$ to a randomly chosen configuration from database without SLO violation. Go to Line~\ref{line:endFor}.
        
        \STATE \textbf{Resource reduction targets:} Determine number of microservice for resource reduction, $n^t$, using Eqn.~\eqref{eq:ns_equation} and resource reduction target for each microservice, $\Delta^t$ using Eqn.~\eqref{eq:ds_equation}.
        
        \STATE \textbf{Avoid bottleneck services:} Get the set $\mathcal{I}^t$ of microservices that do not exceed CPU throttling time threshold.
        
        \STATE \textbf{Microservice-wise augmentation:} Build a new set $\mathcal{I}^{*t}$ from microservices in $\mathcal{I}^t$ with an inclusion probability of $p_i^t$ defined in Eqn.~\eqref{eq:probability}.
        
        \STATE \textbf{Resource reduction:} If {$|\mathcal{I}^{*t}| > n^t$}, uniformly randomly choose $n^t$ microservices from $\mathcal{I}^{*t}$, else choose all microservices from $\mathcal{I}^{*t}$, and then update their resource to $x_i^{t-1}\cdot \Delta^t$.
    \ENDFOR \label{line:endFor}
 \end{algorithmic} 
 \end{algorithm}
 
\begin{figure*}[t!]
\subfigure[Response time vs workload ]{\label{fig:fixed_range_rps_vs_response}\includegraphics[width=0.28\textwidth]{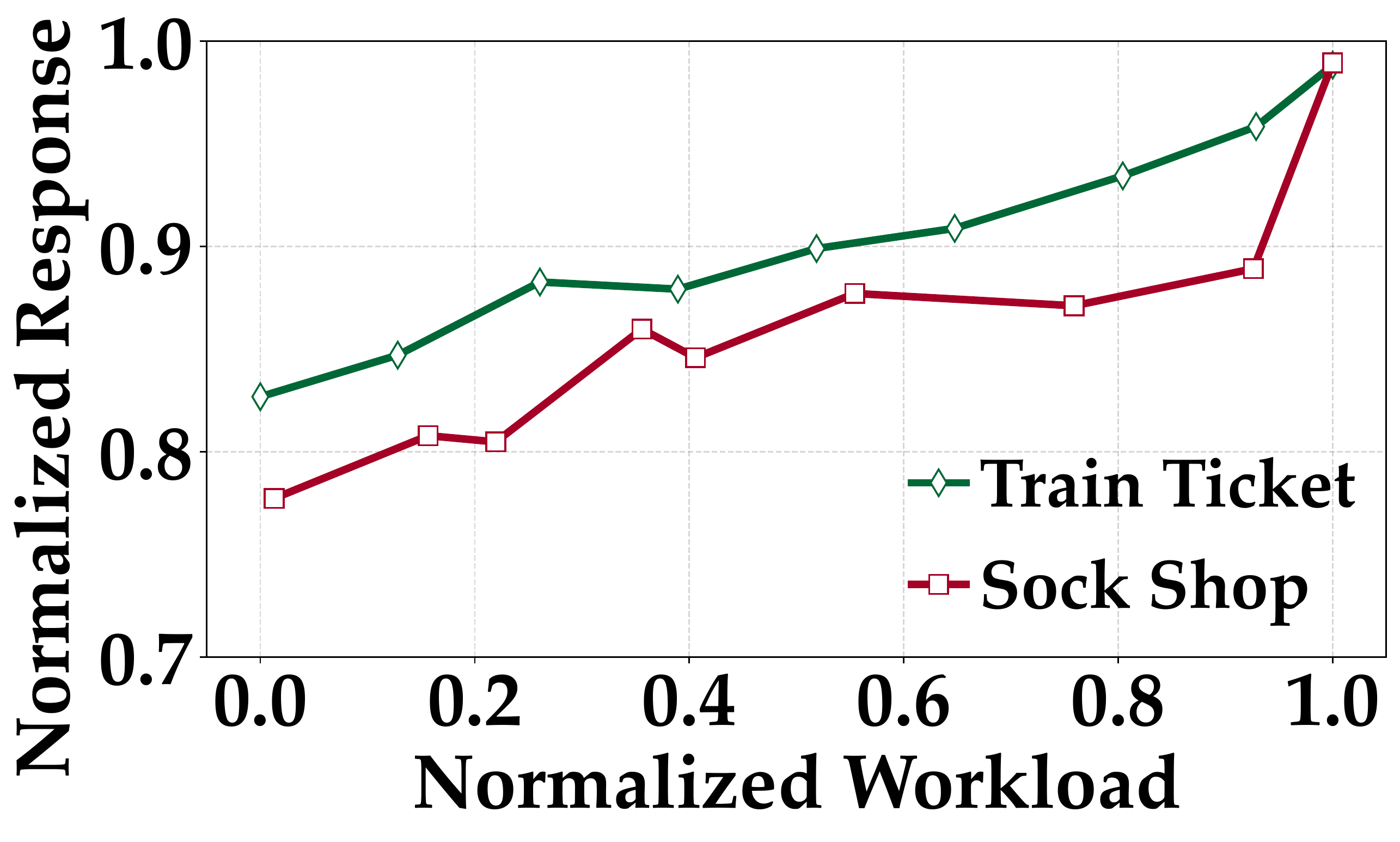}}\hspace{0.1cm}
\subfigure[Dynamic workload range]{\label{fig:database}\includegraphics[width=0.36\textwidth]{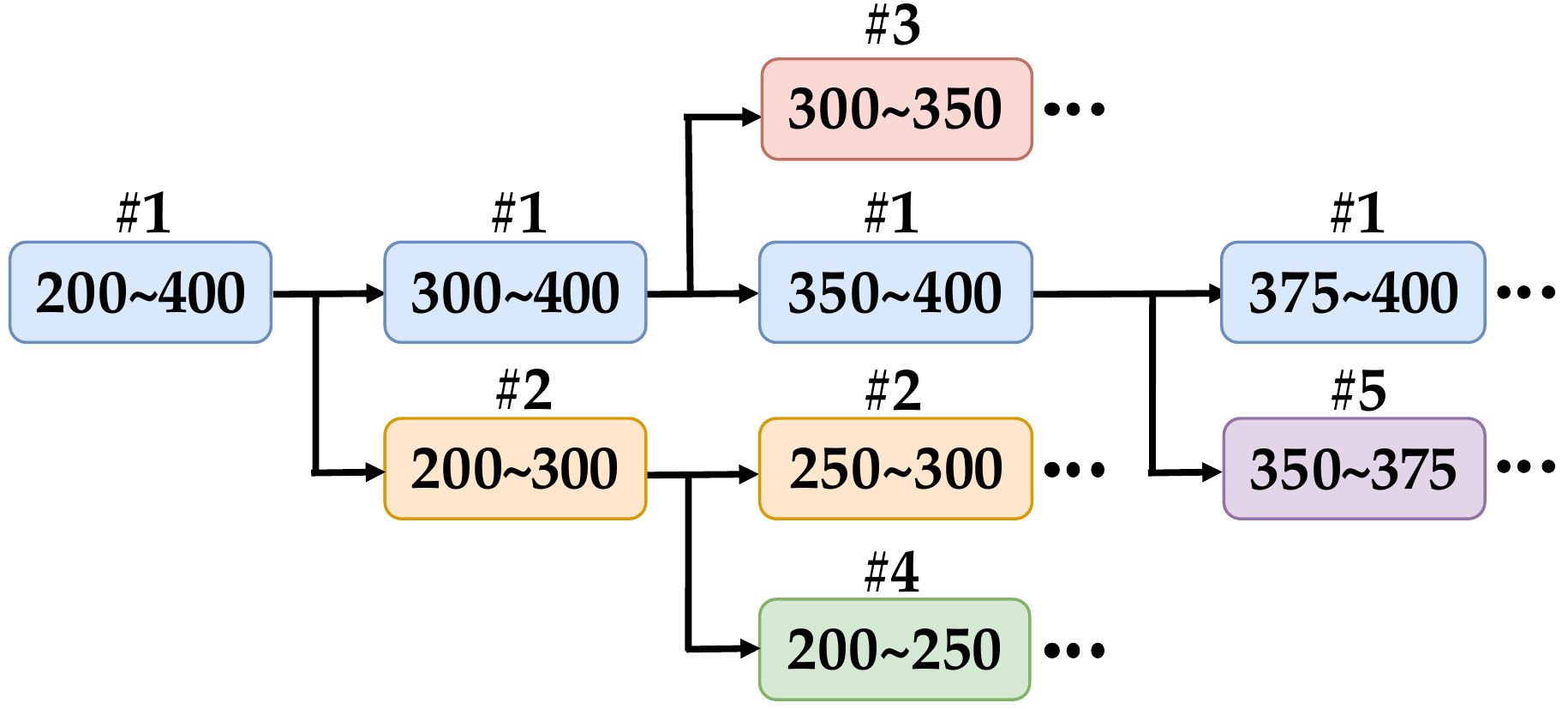}}\hspace{0.1cm}
\subfigure[Dynamic response time target]{\label{fig:dynamic_SLO}\includegraphics[width=0.3\textwidth]{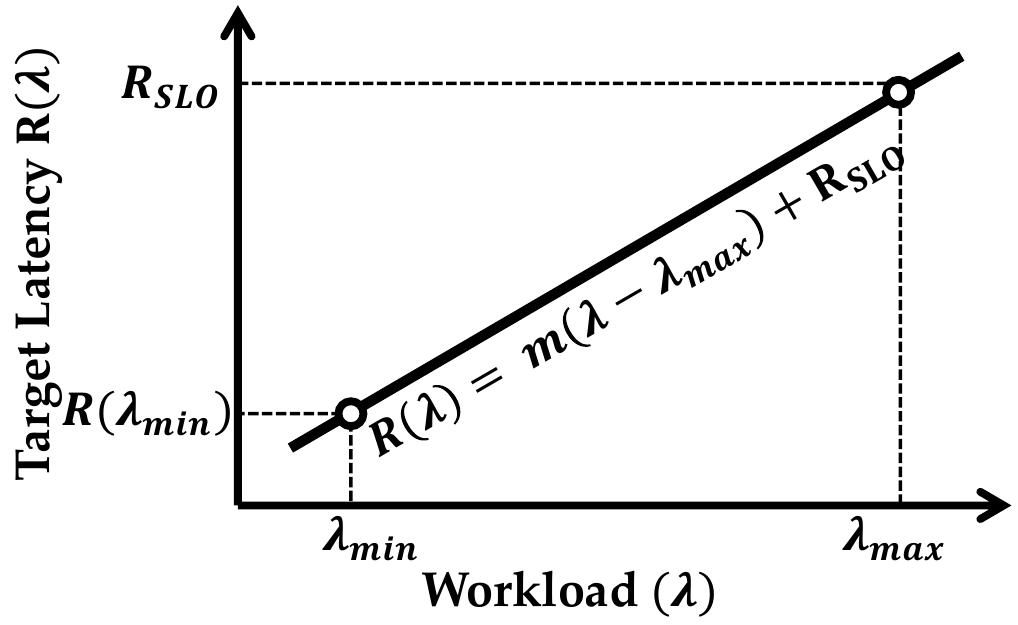}}
\caption{(a) Response time change due to workload. (b) Dynamic workload range to bootstrap efficient resource allocation for different workloads. (c) Dynamically updating target response time to tackle response time change due to workload change.}
\end{figure*}

\textbf{Iterative resource allocation.} \ouralg applies the resource reduction iteratively and saves all resource allocations, $\textbf{x}^t$, and the response times, $r^t$, in a ``resource allocation history database (RHDb)''. \camera{The purpose of the RHDb is to allow \ouralg to roll back to a previous SLO satisfying resource allocation for all microservices in case of an SLO violation.
Even though the resource reduction slows down when the latency approaches the SLO, \ouralg cannot guarantee that its opportunistic resource reduction will never cause an SLO violation.} In addition, changes in microservice implementation or changes in its hardware configuration may also alter optimum resource allocation and cause SLO violations. In such cases, rolling back to a previous configuration allows \ouralg to jump start on finding the new optimum, instead of resetting the resource allocation to the maximum and starting from scratch. \camera{While RHDb itself does not add significant overhead due to its lightweight single-table implementation, the action of rolling back may cause extra iterations for \ouralg to find an efficient resource allocation. Nonetheless, the mechanism of roll back using RHDb is essential for \ouralg's adaptability and QoS assurance.}

\textbf{Escaping sub-optimum configurations.}
The combination of monotonic resource reduction and probabilistic choice of microservices to reduce resource may cause \ouralg to make unfavorable resource reductions early on (e.g., making particular microservice reach bottleneck and push response time close to SLO) and settle at inefficient resource allocation, even though other microservices have redundant resources.
This can force \ouralg to slow-down prematurely, even stop further resource reduction. 
To escape from such inefficient resource allocations, we implement random exploration where \ouralg with a probability $p_e^t$ rolls back to a uniformly random previous resource allocation in RHDb. We set $p_e^t$ based on the response latency as follows
\begin{align}\label{eq:exploration}
    p_e^t =  A \cdot \min\left(\frac{R-r^{t-1}}{\alpha R},1\right)+ B
\end{align}
Here, $A$ and $B$ are exploration parameters that decide the maximum and the minimum probability of exploration, respectively, and satisfy $0\leq B\leq A \leq 1$ and $A+B\leq 1$. The exploration probability decreases as \ouralg's response time $r^{t-1}$ approaches the SLO $R$. The random exploration also allows \ouralg to ``walk back'' the resource reduction path it took and identify previously missed reduction opportunities. Naturally, the degree of exploration affects how quickly we reach an efficient resource allocation. Nonetheless, we do not anticipate this exploration to add significant overhead since \ouralg can find an efficient resource allocation in a few tens of iterations.

\textbf{Implementation of \ouralg.}
We present the working principle in Algorithm~\ref{alg:main_algorithm} where \ouralg takes performance metrics from the system using Prometheus and Linkered and then updates the resource allocation of the microservices while keeping a log of all resource allocations and response times in its database RHDb. The high-level architecture block diagram of \ouralg is presented in Fig.~\ref{fig:pema_system_design}.

\subsection{Workload-Aware Resource Allocation}

Our design of \ouralg so far addresses how we can navigate to find an efficient resource allocation for our microservice-based application.
Our design, through configuration rollback, can also handle changes in microservice implementation. 
Here we address 
how \ouralg tackles the workload variations.
For any cloud application, the workload intensity (i.e., requests per second) directly affects the response time, and hence, how much resource is needed \cite{gandhi2012autoscale,google_autoscale,azure_autoscale}. In Fig.~\ref{fig:fixed_range_rps_vs_response}, we show the change in response time as the workload changes. As \ouralg iteratively makes resource reductions based on the response time, a decrease in workload will falsely indicate resource reduction opportunities that do not work for high workloads, leading to many SLO violations when the workload increases. The same is true for prior ML-based approaches that do not explicitly address workload change \cite{gan2021sage,gan2019seer}.

Hence, \ouralg needs to identify efficient resource allocations at different workload levels. A straightforward way is to divide the workload variations into discrete workload ranges \camera{(e.g., a workload range from ``X'' requests-per-second to ``Y'' requests-per-second)} and run multiple copies of \ouralg in a ``pseudo-parallel'' fashion. We say pseudo-parallel as at any time only one \ouralg is working on its corresponding workload range.
Note here that the workload ranges need to be small enough to not significantly affect the response latency, requiring resource allocation changes, i.e., a single resource allocation should work for the entire range. For instance, a range of 25 requests-per-second in \trainT microservice is a suitable workload range.

\textbf{Dynamic workload-range.}
While in principle multiple parallel \ouralg works, it may take a long time to reach efficient allocations for every workload range. To accelerate the learning, we propose a novel approach where we start with a few (two/three) larger workload ranges and gradually split each range (i.e., parent range) into smaller ranges (i.e., child range) until we reach our target workload ranges. The goal here is to utilize learning from the parent ranges to bootstrap the learning process for the child ranges. During a range split, the parent range is divided into two equal child ranges. We attach \ouralg of the parent range to the child range with a higher workload, whereas a new \ouralg process is launched for the other child range. The new \ouralg uses the resource allocations of the parent range as the starting point and requires fewer iterations to reach an efficient resource level.
The intuition for this approach is that a resource allocation that satisfies SLO at a higher workload should also satisfy SLO for a lower workload. Fig.~\ref{fig:database} illustrates the idea where we start with a workload range of 200$\sim$400 and then branch out to smaller ranges. The number on top of each range identifies the \ouralg process attached to this range. The original \ouralg process with id ``\#1'' remains attached to the higher workload ranges (e.g., 300$\sim$400, 350$\sim$400, 375$\sim$400) as we split each range into smaller ranges.

\textbf{Dynamic response time target.}
While this approach benefits the learning time, we need to tackle the latency variation due to workload changes when the workload ranges are large (e.g., 200$\sim$400 rps for \trainT). We use one \ouralg process for each workload range, even during the initial stages with large ranges (e.g., \ouralg \#1 for 300$\sim$400 range in Fig.~\ref{fig:database}). Each \ouralg process needs to make an SLO preserving resource allocation that works for its entire range. 
To achieve this, instead of setting it to the SLO specificity response time, we update $R$ in Eqns.~\eqref{eq:ns_equation}, \eqref{eq:ds_equation}, and \eqref{eq:exploration} into a function of workload $\lambda$ as follows
\begin{align} \label{eq:dynamic_SLO}
    R(\lambda) = m \cdot (\lambda - \lambda_{max})+R_{SLO}
\end{align}
Here, $m$ is a parameter that determines the change in latency performance for a unit change in workload, $\lambda_{max}$ is the upper limit of a workload range, and $R_{SLO}$ is the SLO specified response time.
Fig.~\ref{fig:dynamic_SLO} illustrates our approach of using a dynamic response time target. We see from Eqn.~\eqref{eq:dynamic_SLO} that when the workload is low within a range, we set a conservative (i.e., lower than SLO) latency target to intentionally allocate more resource than needed and therefore allow headroom for higher workloads. This approach intentionally makes conservative inefficient resource allocations for lower workload levels within a range. However, as the ranges get smaller as we split them, the latency variation within a range also gets smaller, and so is the inefficiency. On the other hand, we learn $m$ at the beginning of \ouralg when we keep the resource allocation fixed for a few time steps while the workload changes. We then use linear regression on the workload vs response time (as in Fig.~\ref{fig:fixed_range_rps_vs_response}) to extract $m$.
Note that we learn $m$ only once at the beginning when the workload ranges are large. During range splits, we keep the $m$ from the parent range. Now, $m$ may change as we make the resource allocations change on the microservice. Nonetheless, as our range split reaches the final workload ranges, we no longer need the dynamic response target, and $m$ becomes irrelevant.

\subsection{Handling Transient Events}
From our extended experiments we identify that \ouralg is susceptible to unnecessary SLO violations due to transient dips in the response time. More specifically, after \ouralg has already identified an efficient allocation, a momentary/transient dip in response time drives \ouralg to make resource reductions only to meet with SLO violation in the next iteration. To circumvent this, we adopt a moving average approach where we take the average of the response time of $K$ recent time steps and update the $n^t$ and $\Delta^t$ as follows
\begin{align}
  n^t &= N\cdot \min \left( \frac{R-\frac{1}{K}\sum_{k=1}^K r^{t-k}}{\alpha R},1\right)\\
  \Delta^t &= \beta \cdot \min \left( \frac{R-\frac{1}{K}\sum_{k=1}^K r^{t-k}}{\alpha R},1\right) \cdot 100\%
\end{align}
Note that, to ensure QoS, we do not apply this moving averaging for detecting SLO violations. We still roll back resource allocations based on the most recent response time as in Line 4 in Algorithm~\ref{alg:main_algorithm}.

\section{Evaluation}
We use our microservice application prototypes, \trainT, \socS, and \hotelR, to evaluate \ouralg.
Here we first discuss details of \ouralg's execution followed by performance evaluation against other resource allocation strategies. We then present how different parameters affect \ouralg, and finally show how \ouralg can adapt to change in operating conditions.

\begin{figure}[t]
\subfigure[CPU allocation]{\label{fig:ss_convergence_2}\includegraphics[width=0.23\textwidth]{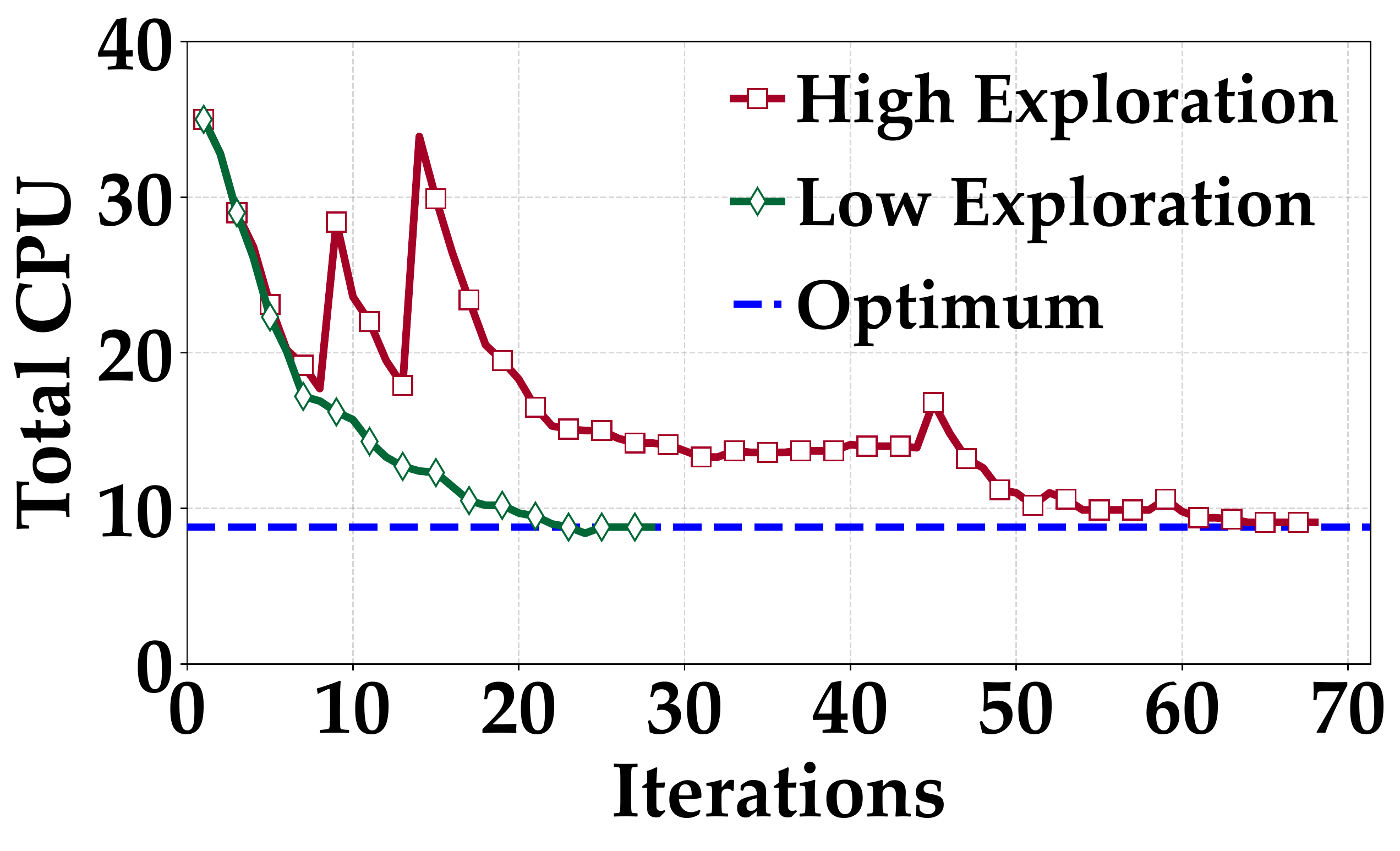}}
\subfigure[Response time]{\label{fig:ss_convergence_response} \includegraphics[width=0.23\textwidth]{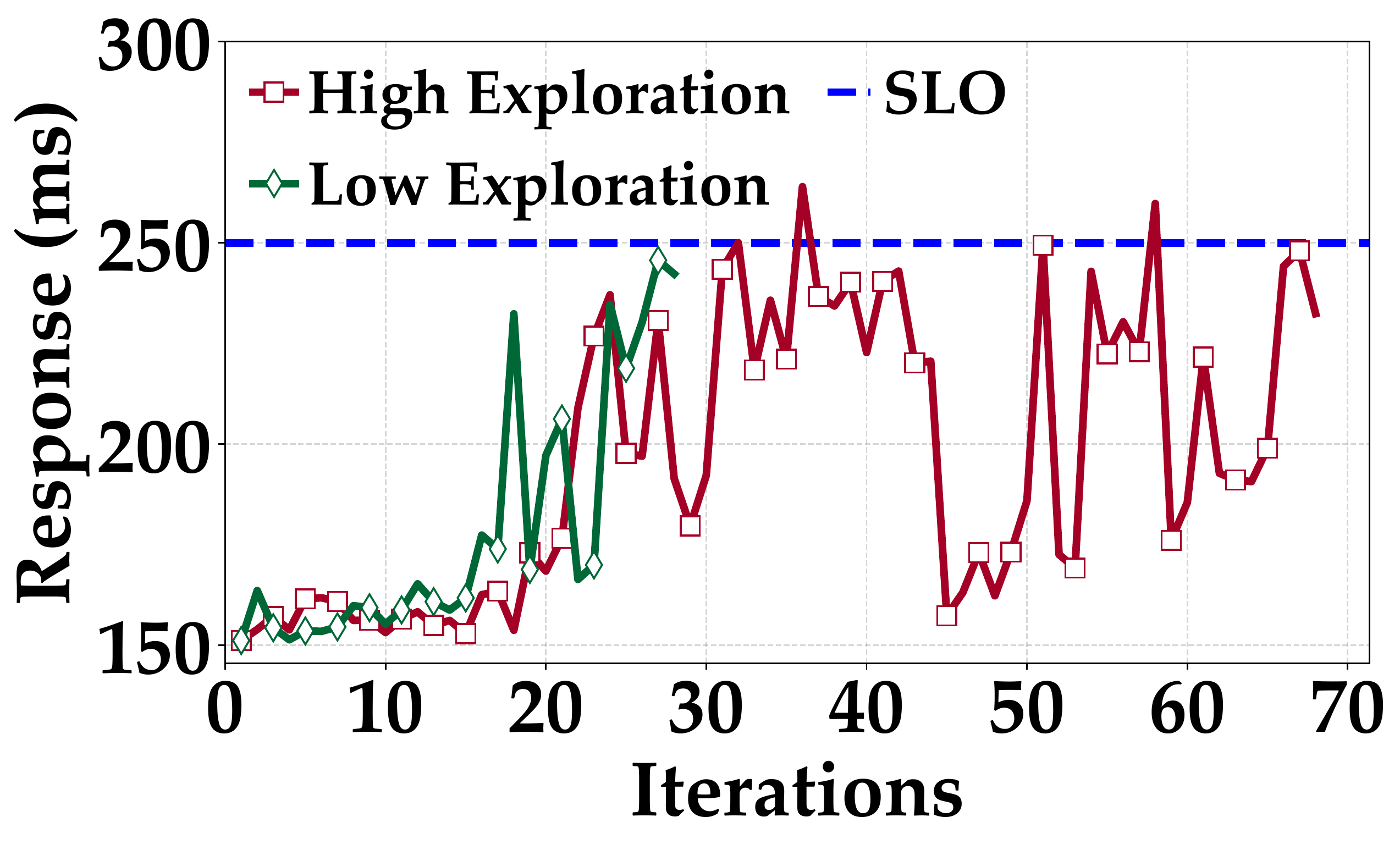}}
\caption{Execution of \ouralg on \socS with different explorations. The exploration parameters in Eqn.~\eqref{eq:exploration} for high exploration are $A = 0.1, B = 0.01$, and for low exploration are $A = 0.05, B = 0.005$.
}
\label{fig:ss_convergence_exp}
\end{figure}

\begin{figure}[t]
\subfigure[\trainT]{\label{fig:ts_cpu_response_convergence}\includegraphics[width=0.23\textwidth]{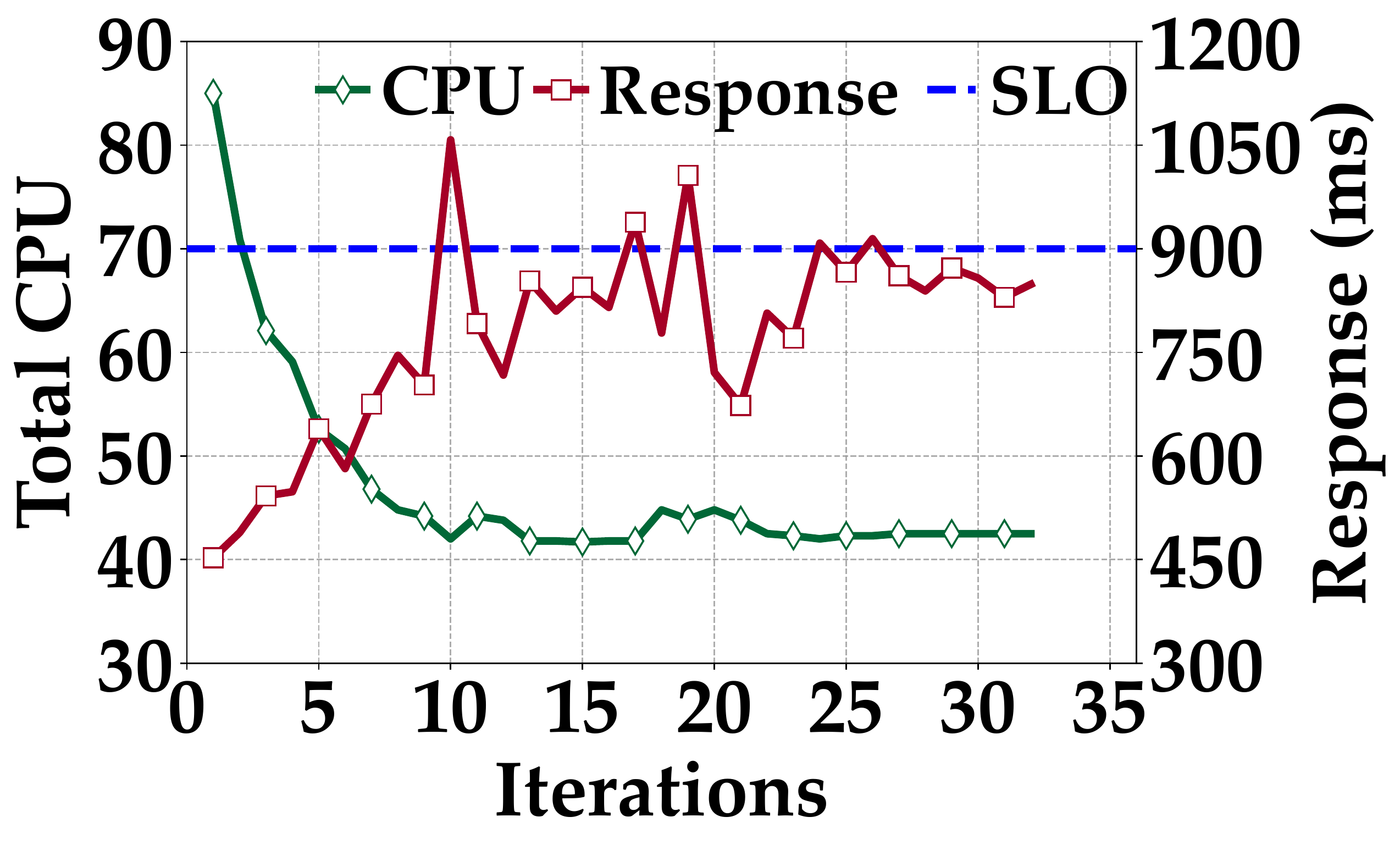}}\hspace{0.2cm}
\subfigure[\hotelR]{\label{fig:hr_cpu_response_convergence}\includegraphics[width=0.23\textwidth]{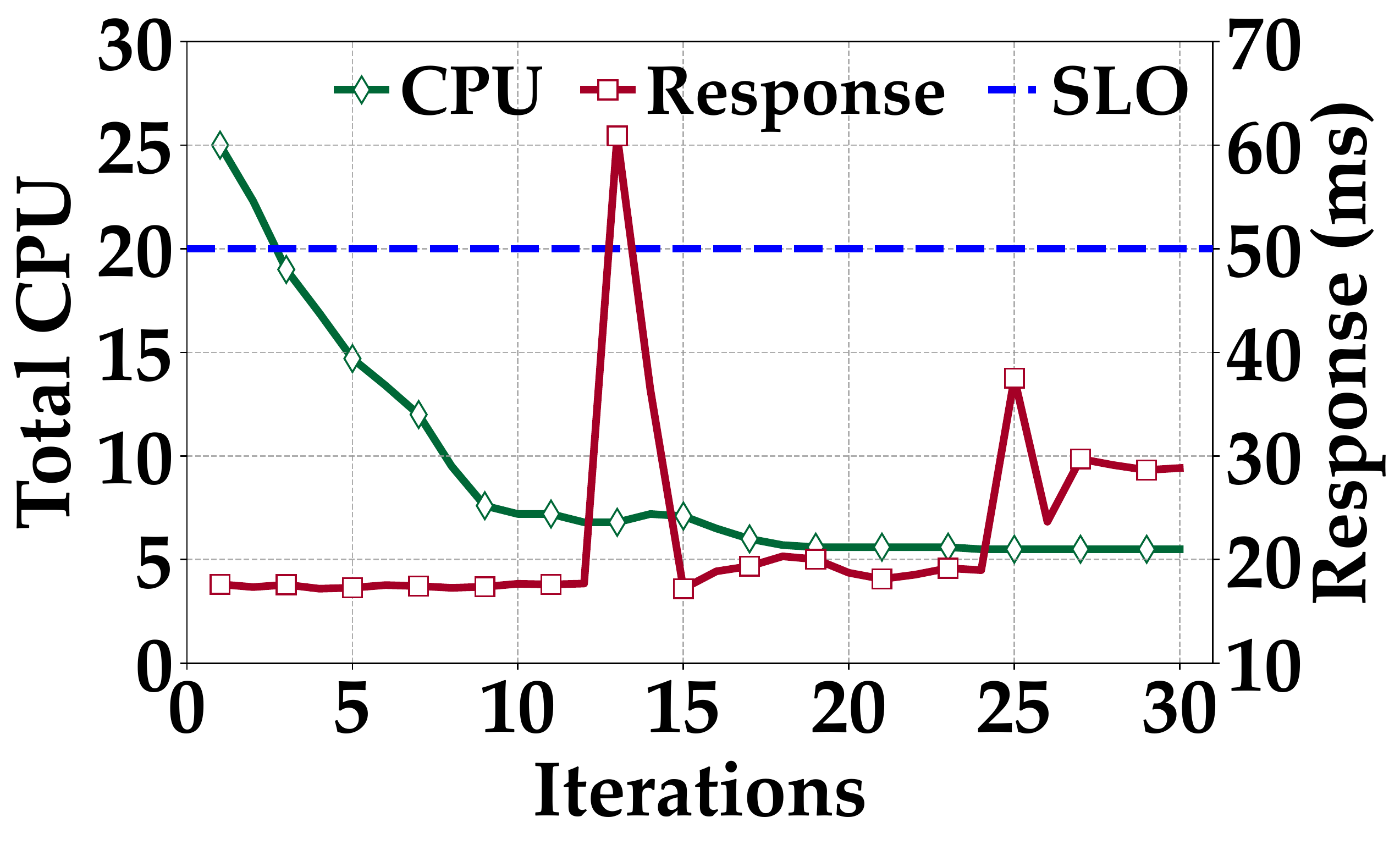}}
\caption{Execution of \ouralg for \trainT and \hotelR.}
\end{figure}

\subsection{Execution of \ouralg}

Here, we first show how \ouralg finds efficient resource allocation using iterative resource reduction, \camera{where the duration of each iteration is two minutes.} We then demonstrate how workload-aware \ouralg utilizes the dynamic workload range and response time target. Finally, we present a 36-hour long experiment with \ouralg making efficient resource allocation maintaining QoS. 

\textbf{Efficient resource allocation.}
Fig.~\ref{fig:ss_convergence_2} demonstrates the iterative resource allocation and Fig.~\ref{fig:ss_convergence_response} shows the corresponding response times for \socS under a workload of 700 requests per second for two different sets of exploration parameters.
\camera{Here, the optimum total CPU allocation is 8.8 which is identified using extensive trial and error.}

We see in Fig.~\ref{fig:ss_convergence_2} that when a higher exploration is used, \ouralg intentionally increases the resource allocation twice around iteration 10 by going back to an older and higher CPU allocation. We also see that \ouralg with high exploration settles at an inefficient allocation after 20 iterations as the response time reaches SLO (Fig.~\ref{fig:ss_convergence_response}). However, due to exploitation, we see that around iteration 45, it rolls back to an older allocation and finds its way to the efficient allocation. Incidentally, \ouralg with low exploration also reaches the efficient resource allocation. We see a few SLO violations in Fig.~\ref{fig:ss_convergence_response} which are mitigated immediately by increasing the CPU resource. Figs.~\ref{fig:ts_cpu_response_convergence} and \ref{fig:hr_cpu_response_convergence} show the iterative resource change and the corresponding response times for \trainT and \hotelR, respectively. 

Regardless of the microservice implementation, we see that \ouralg can successfully find efficient resource allocations with only a few unintentional SLO violations.

\begin{figure}[t!]
\subfigure[Total CPU allocation over different ranges]{\label{fig:dynamic_range_CPU}\includegraphics[width=0.23\textwidth]{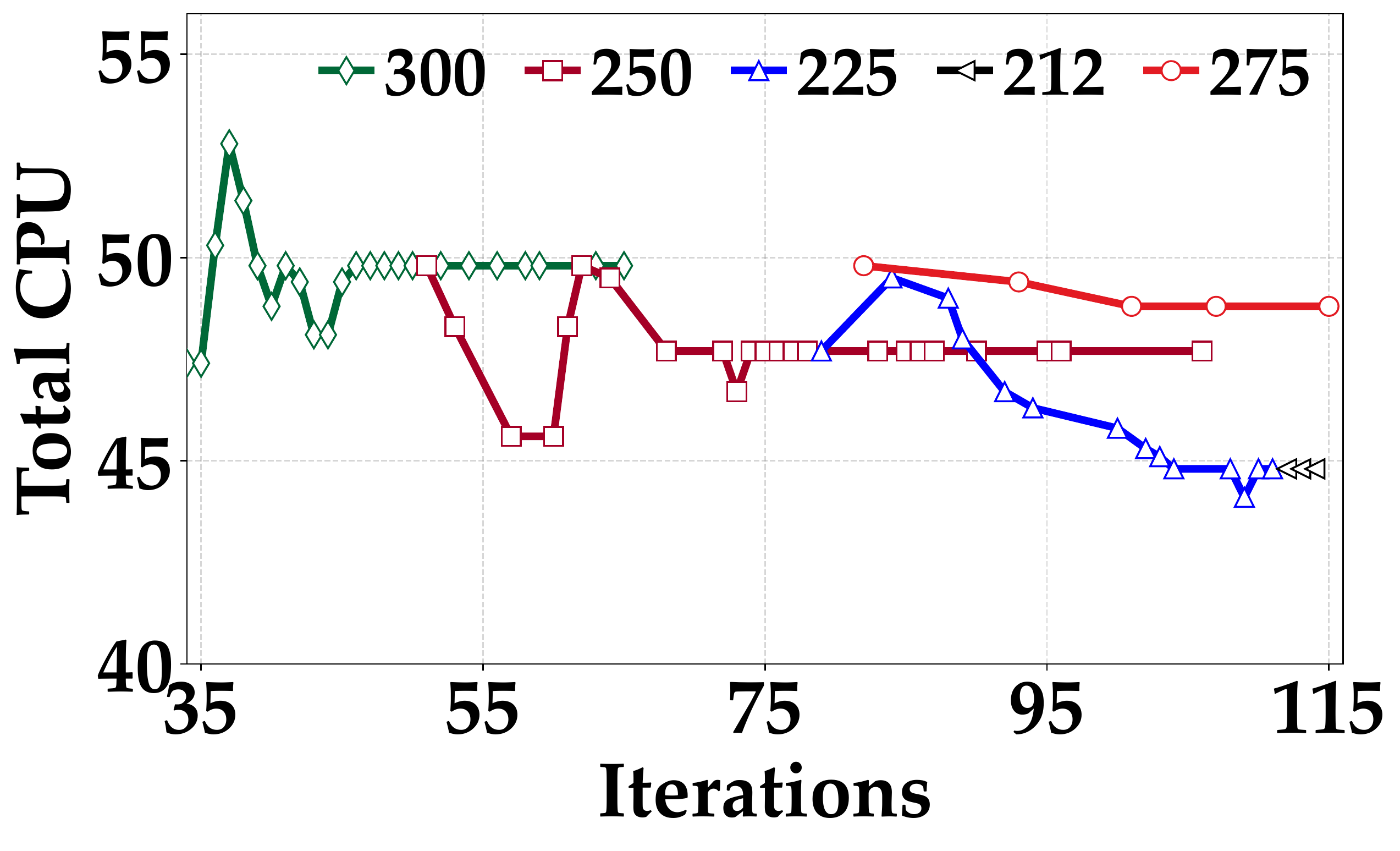}}
\subfigure[Response time of the iterations]{\label{fig:dynamic_range_response}\includegraphics[width=0.23\textwidth]{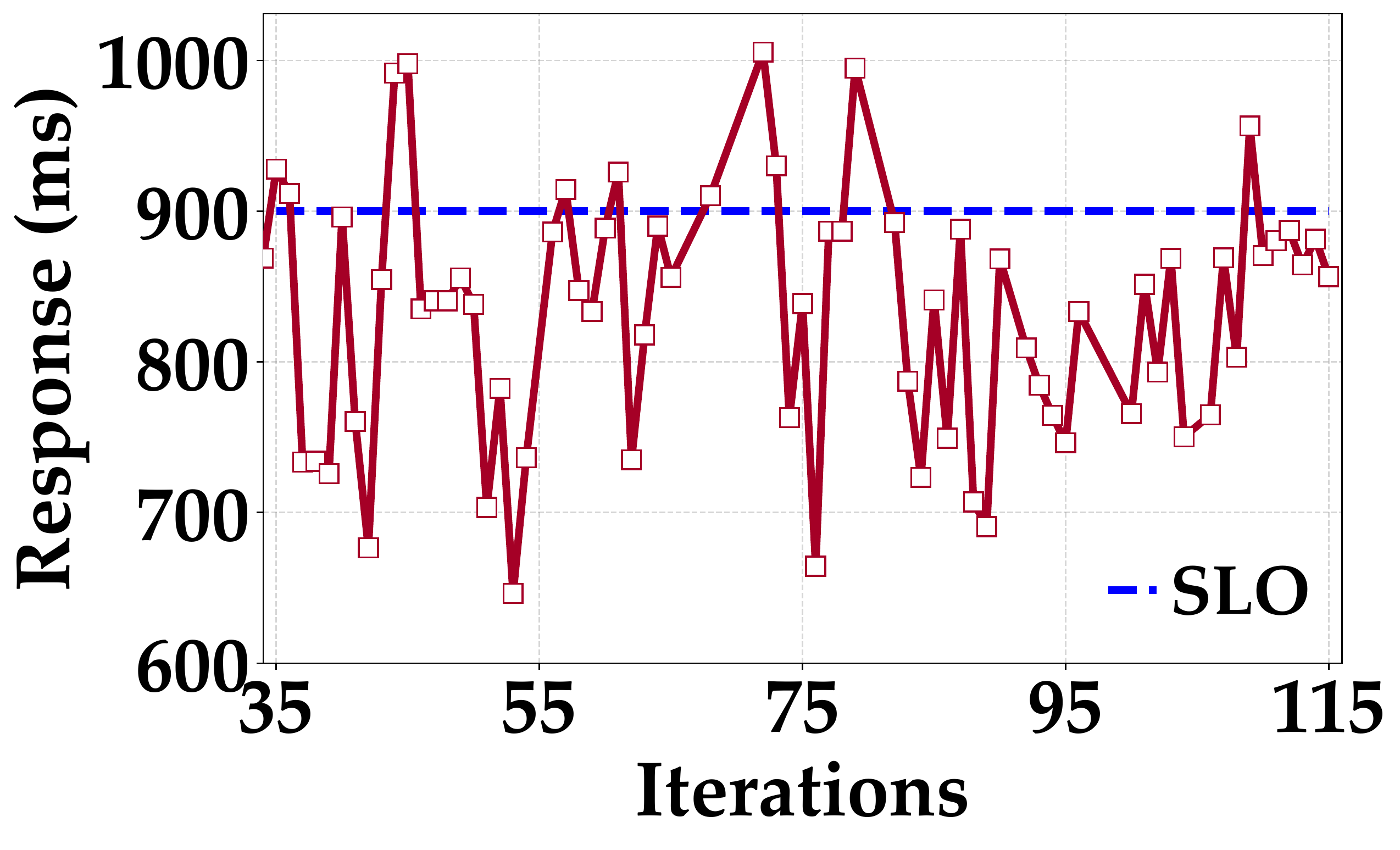}}
\caption{Execution of \ouralg on \trainT with dynamic workload range. (a) CPU allocation. (b) Response time.}
\label{fig:ts_automatic_range_identification}
\end{figure}

\begin{figure}[t!]
\subfigure[]{\label{fig:workload_trace}\includegraphics[width=0.48\textwidth]{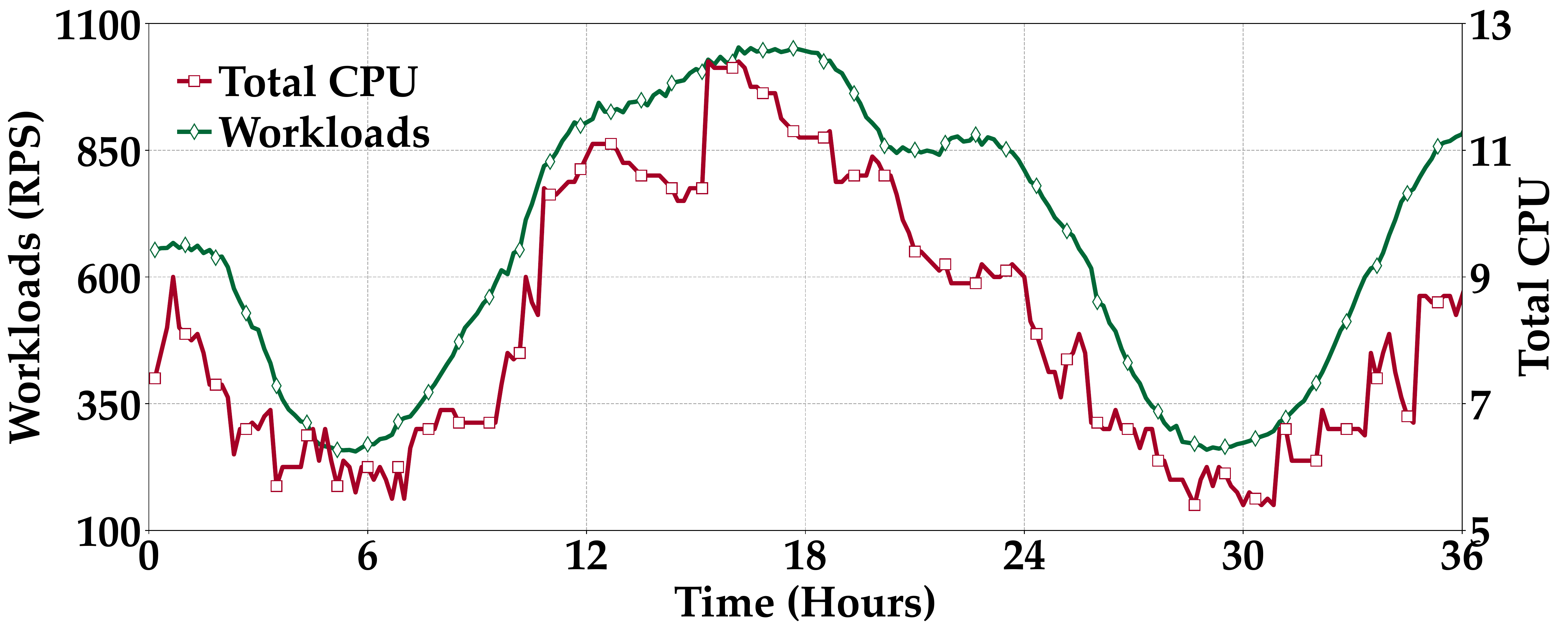}}
\subfigure[]{\label{fig:long_response}\includegraphics[width=0.48\textwidth]{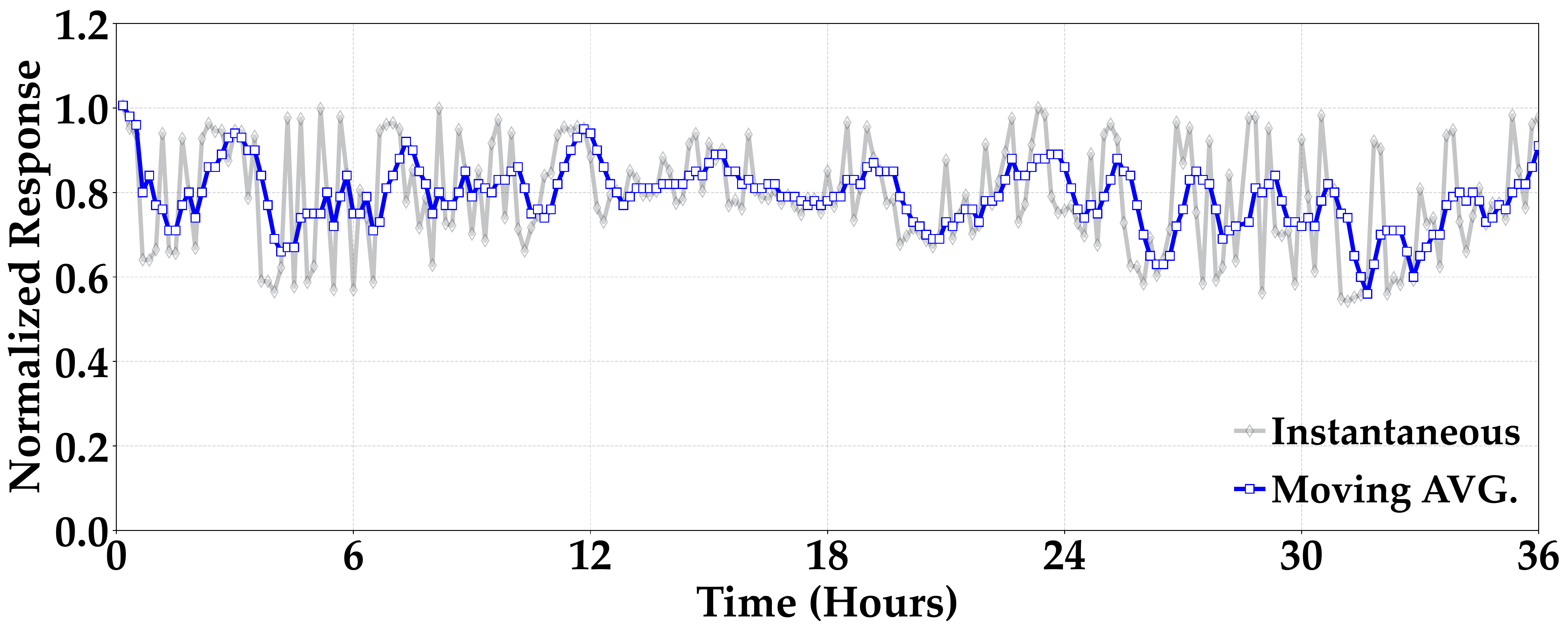}}
\caption{Extended execution of \ouralg in \socS. (a) Workload and CPU allocation. (b) Response time normalized to SLO.}
\label{fig:ss_1_days_trace}
\end{figure}

\textbf{Dynamic workload range.}
Next, in Fig.~\ref{fig:dynamic_range_CPU}, we show the resource allocation of \ouralg for \trainT as our workload varies between 200 and 300 requests per second. The legend in this figure indicates the upper limit on the workload range. The workload range 300 (i.e., 200$\sim$300) first splits into ranges 300 and 250 around iteration 50. The 250 range splits into 250 and 225 around iteration 80, while the 300 splits into 300 and 275 right before iteration 85. We see that each workload range finds an efficient allocation within a few iterations as they start from an already good allocation. Fig.~\ref{fig:dynamic_range_response} shows the corresponding the response time. We see some SLO violations, which are mitigated by \ouralg.

\textbf{Extended execution.}
We run \ouralg on \socS for 36-hour where we change the workloads between 200 and 1100 requests per second following the workload pattern of Wikipedia collected from \cite{wiki_trace_data}. Fig.~\ref{fig:workload_trace} shows the workload pattern and the corresponding resource allocation. We see that \ouralg varies the total resource allocation with changing workload to maintain efficient allocation. Note here that simply varying scaling resource allocation based on workload does not work on microservices as the distribution of the resource plays an important role in performance. Fig.~\ref{fig:long_response} shows the corresponding response times. We show both the instantaneous (i.e., most recent) and moving average responses with a window size of five. Recall that \ouralg reduces resources based on the moving average to avoid transient changes while tackling SLO violation based on the instantaneous response time.

\begin{figure}[t!]
\subfigure[\trainT]{\label{fig:ts_performance_comparisons}\includegraphics[width=0.32\linewidth]{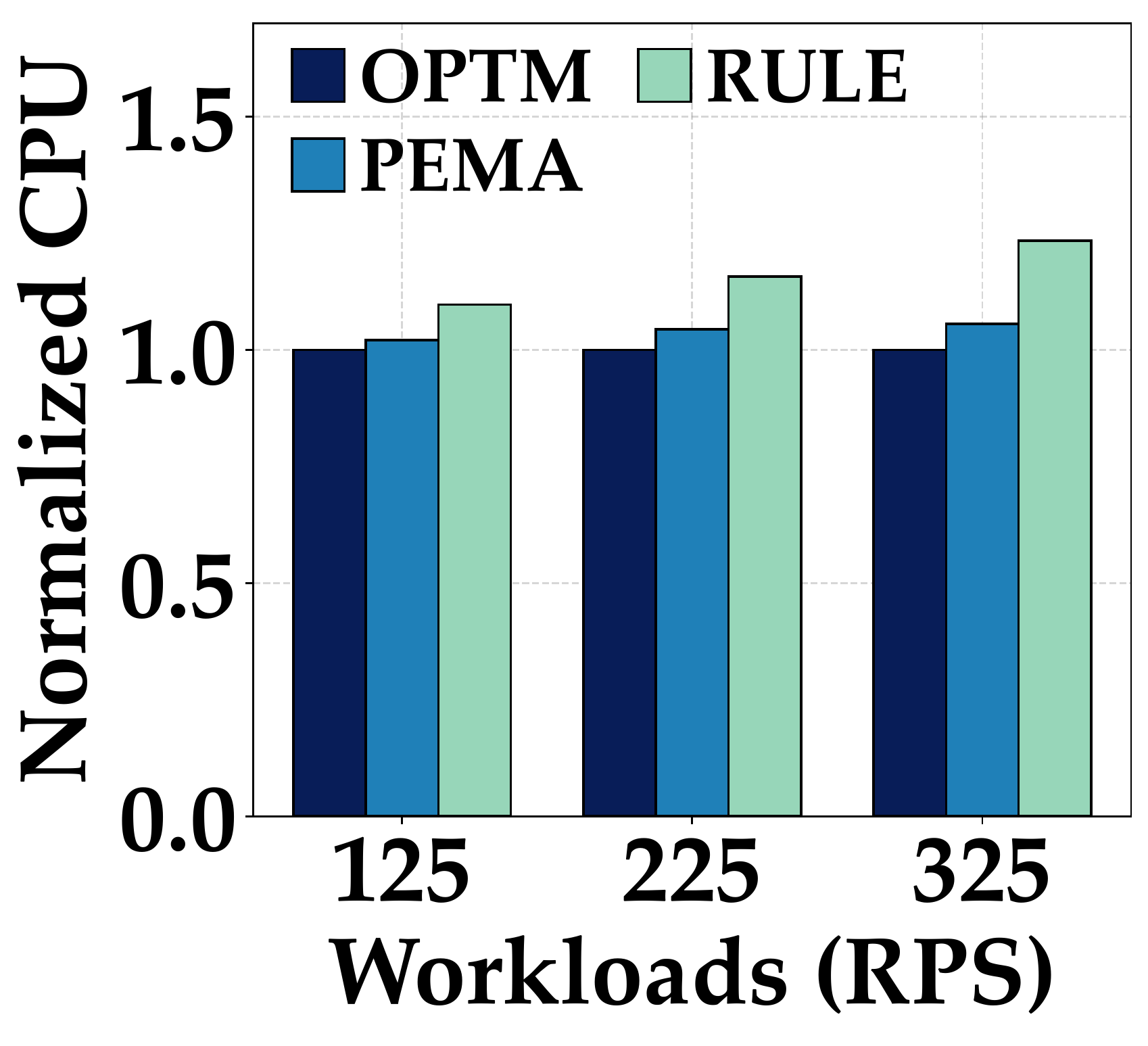}}
\subfigure[\socS]{\label{fig:ss_performance_comparisons}\includegraphics[width=0.32\linewidth]{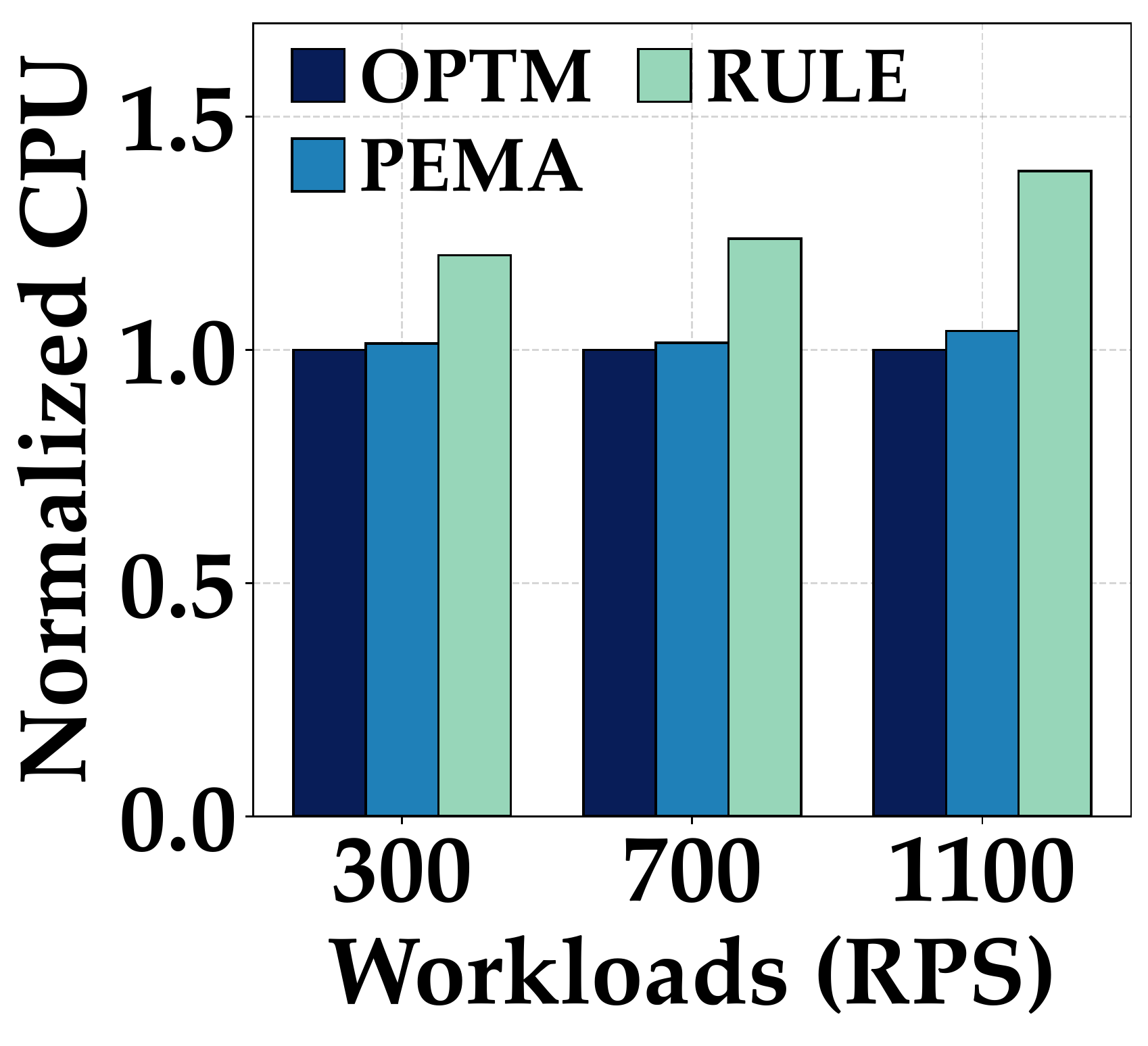}}
\subfigure[\hotelR]{\label{fig:hr_performance_comparisons}\includegraphics[width=0.32\linewidth]{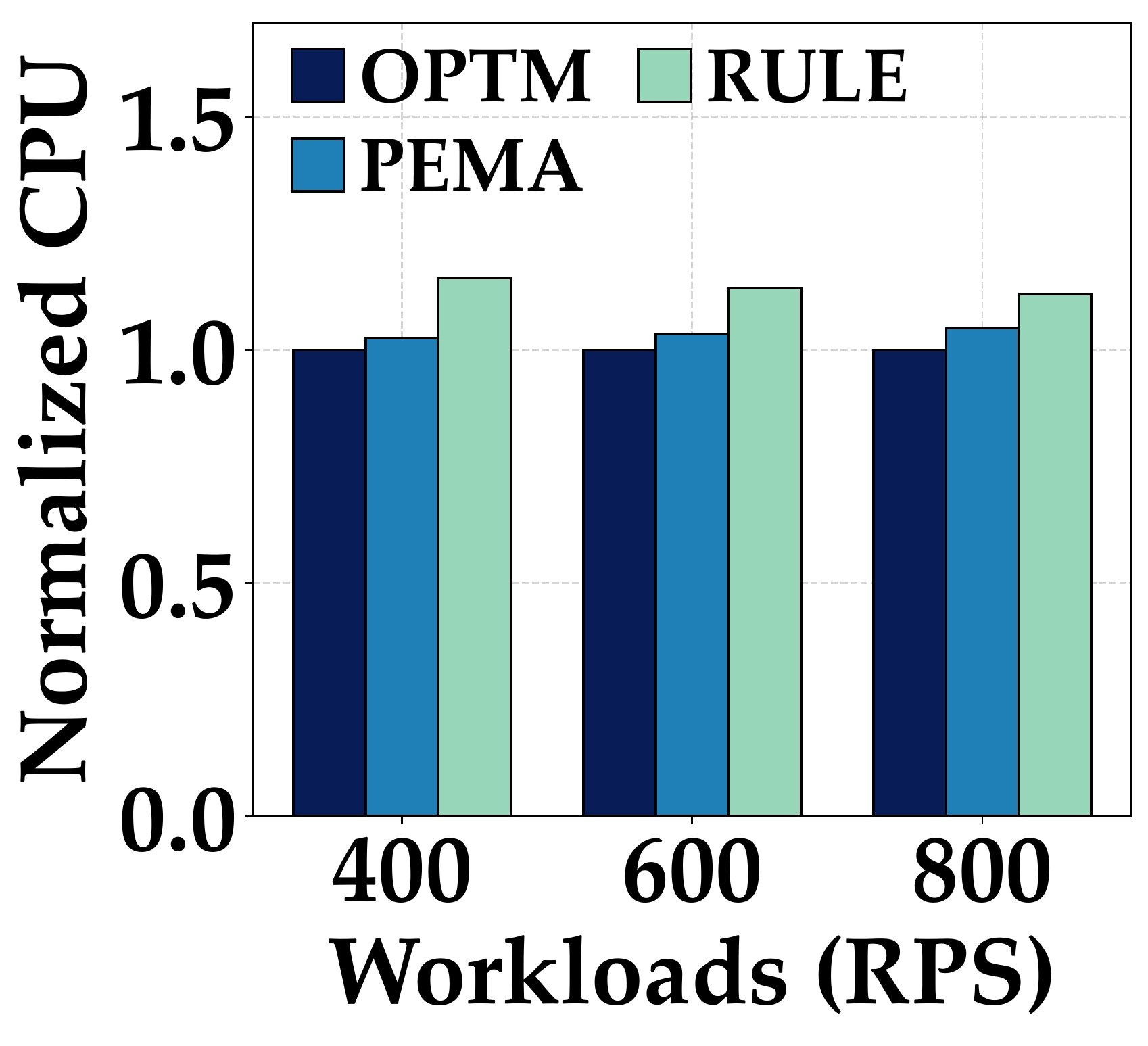}}
\caption{Performance comparison of \ouralg against optimum (\opt) and commercial autoscaler (\ruleA). The CPU allocation is normalized to that of \opt. \ouralg is close to optimum and saves up to 33\% resource compared to \ruleA.}
\vspace{-0.6cm}
\end{figure}

\subsection{Performance evaluation} \label{sec:benhmark}
\textbf{Benchmark strategies.}
We compare the resource allocation efficiency of \ouralg against two benchmark strategies - optimum (\opt) and rule-based (\ruleA). In \opt, we use an exhaustive trial and error search to identify the best possible resource allocation. \camera{We identify a resource allocation as optimum if a small resource reduction (in our case 0.1 CPU) in any of the microservices results in a SLO violation. Note that, \opt cannot be used in practice as it causes many SLO violations during trial and error.} It acts as the upper limit of resource efficiency achievable by any resource manager. \ruleA is Kubernetes' rule-based resource scaling \cite{kubernetes_hpa}. We chose \ruleA as a commercially available resource allocation algorithm to gauge \ouralg's efficiency improvement. We do not compare \ouralg to the ML-based resource allocation strategies as they do not focus on resource allocation efficiency.

\textbf{Comparison of resource allocation efficiency.}
We run each of the three microservices applications using \ouralg and the two benchmark algorithms. Since \opt requires extensive manual search, we evaluate these algorithms for three different workload levels for each microservice. Also, since \ouralg is provably efficient, we run \ouralg several times under each setting and show the average resource allocation.
We normalize each resource allocation for each workload level using the resource allocation of \opt.

Figs.~\ref{fig:ts_performance_comparisons}, \ref{fig:ss_performance_comparisons}, and \ref{fig:hr_performance_comparisons} show the resource allocations of \trainT, \socS, and \hotelR, respectively for the three different algorithms. We see that \ouralg's resource allocation efficiency is very close to \opt. We also observe that \ouralg's efficiency drifts away with increasing workload. On the other hand, \ouralg consistently beats \ruleA, saving as much as 33\% on resource allocation for \socS at high workloads.

The performance comparison results demonstrate that despite being a lightweight resource manager, \ouralg can deliver close to optimum resource allocation while retaining its capability to tackle workload variation without any significant overhead (e.g., ML training).

\begin{figure}
\subfigure[Resource allocation]{\label{fig:combine_variable_alpha_fixed_beta_resource}\includegraphics[width=0.23\textwidth]{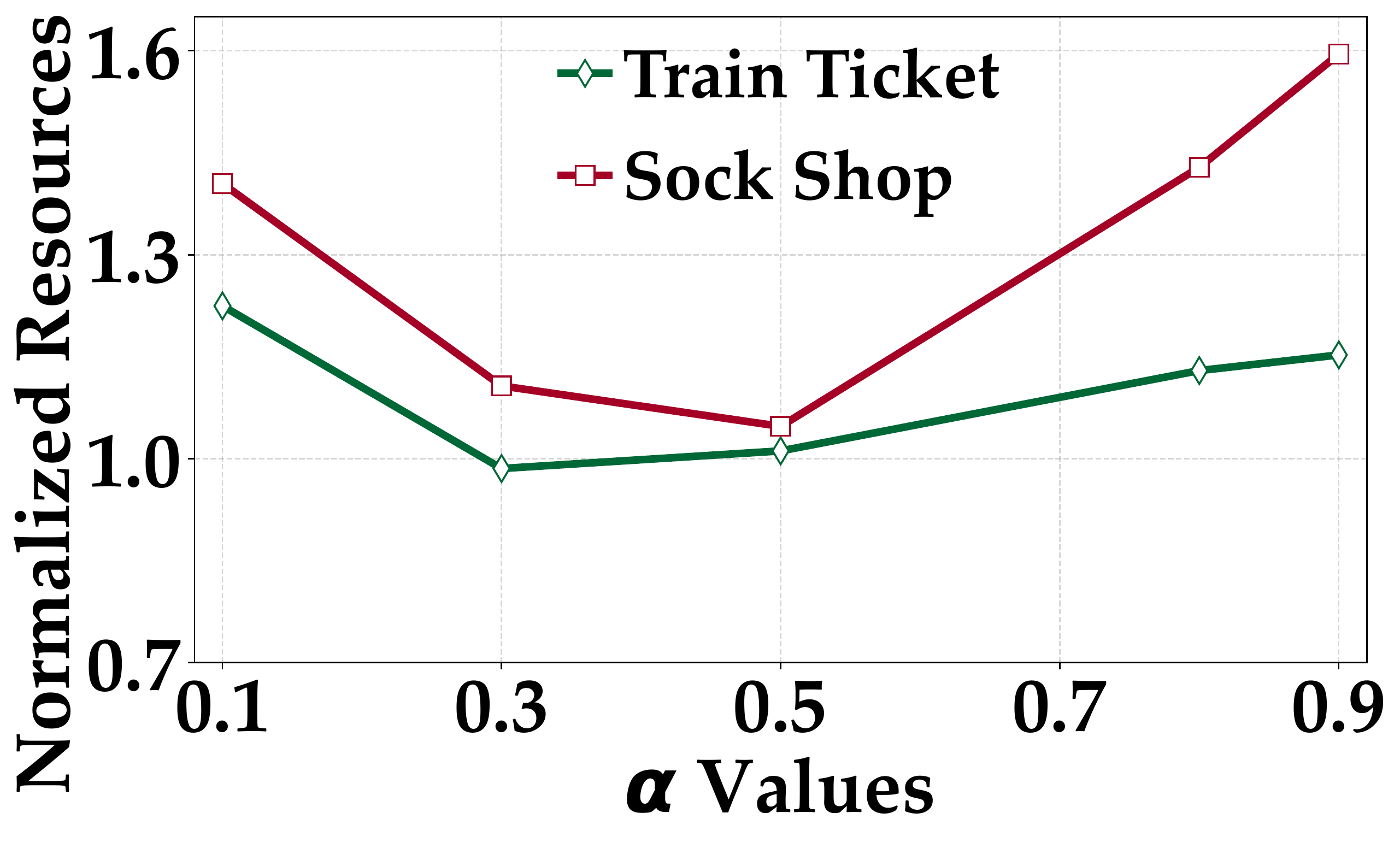}}\hspace{0.2cm}
\subfigure[SLO violations]{\label{fig:combine_variable_alpha_fixed_beta_slo_violations}\includegraphics[width=0.23\textwidth]{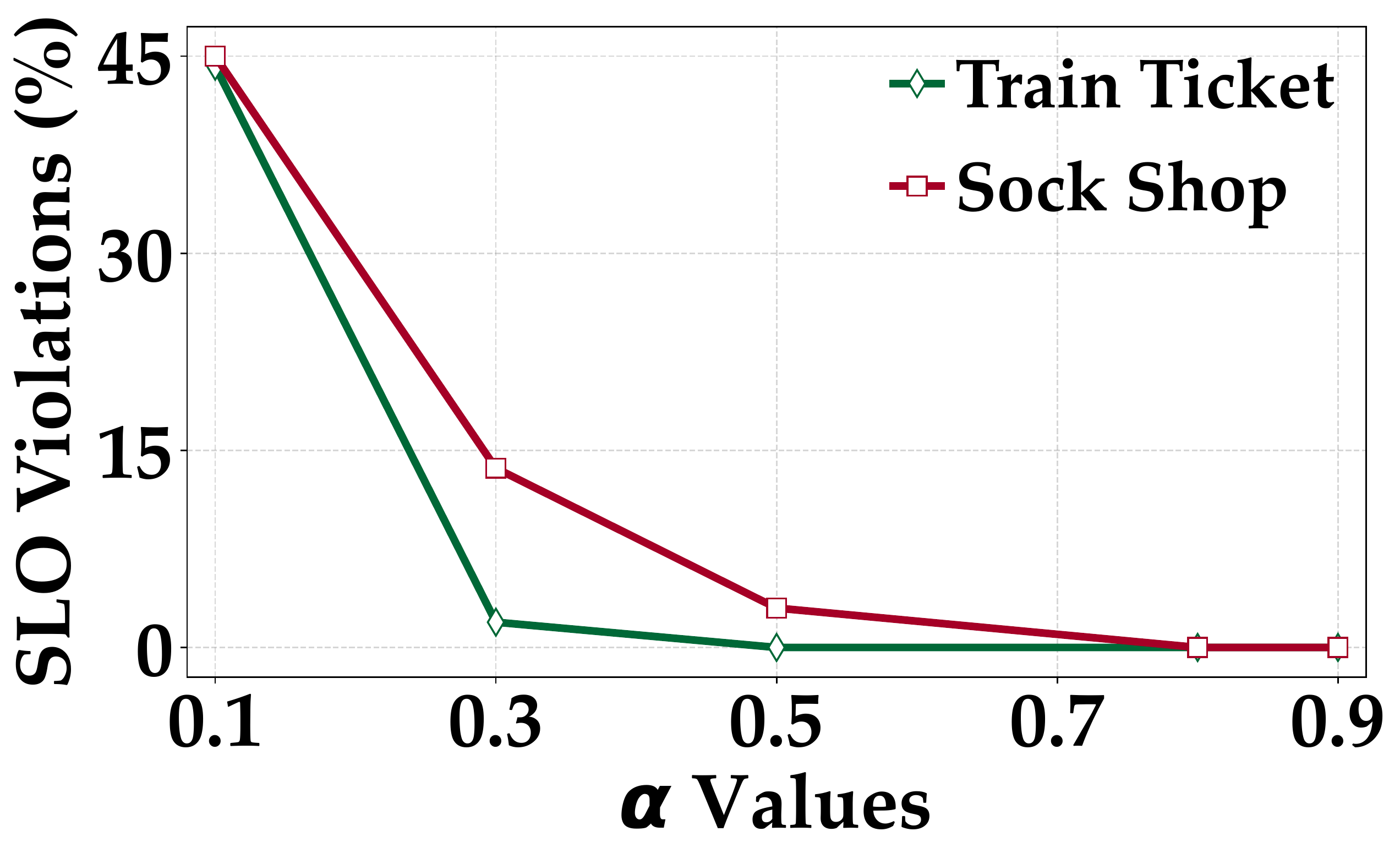}}
\caption{\ouralg's sensitivity to $\alpha$ for a $\beta=0.3$ (a) Resource allocation normalized to optimum. (b) SLO violations.}
\end{figure}

\begin{figure}
\subfigure[Resource allocation]{\label{fig:beta_CPU}\includegraphics[width=0.23\textwidth]{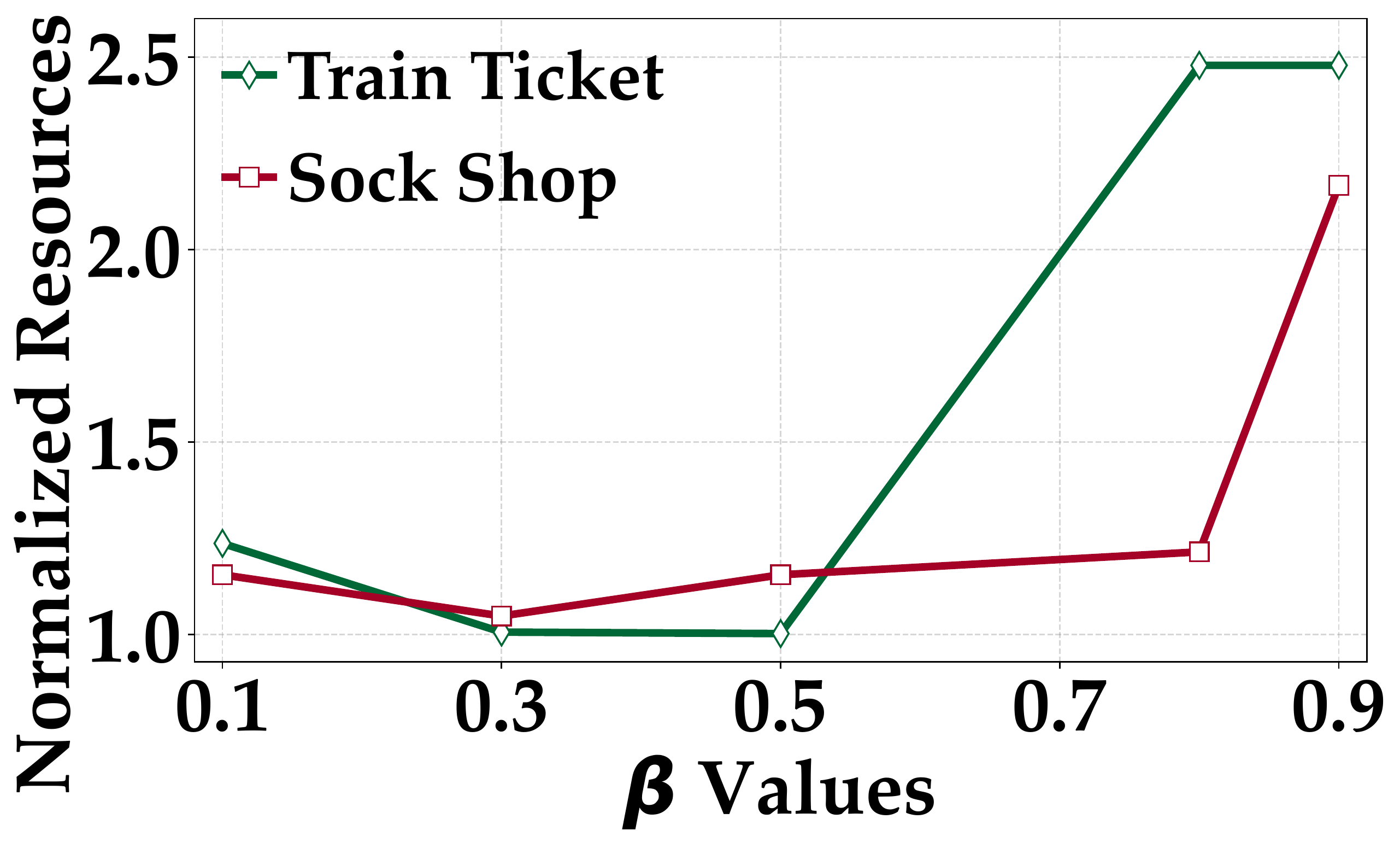}} \hspace{0.1cm}
\subfigure[SLO violations]{\label{fig:beta_SLO}\includegraphics[width=0.23\textwidth]{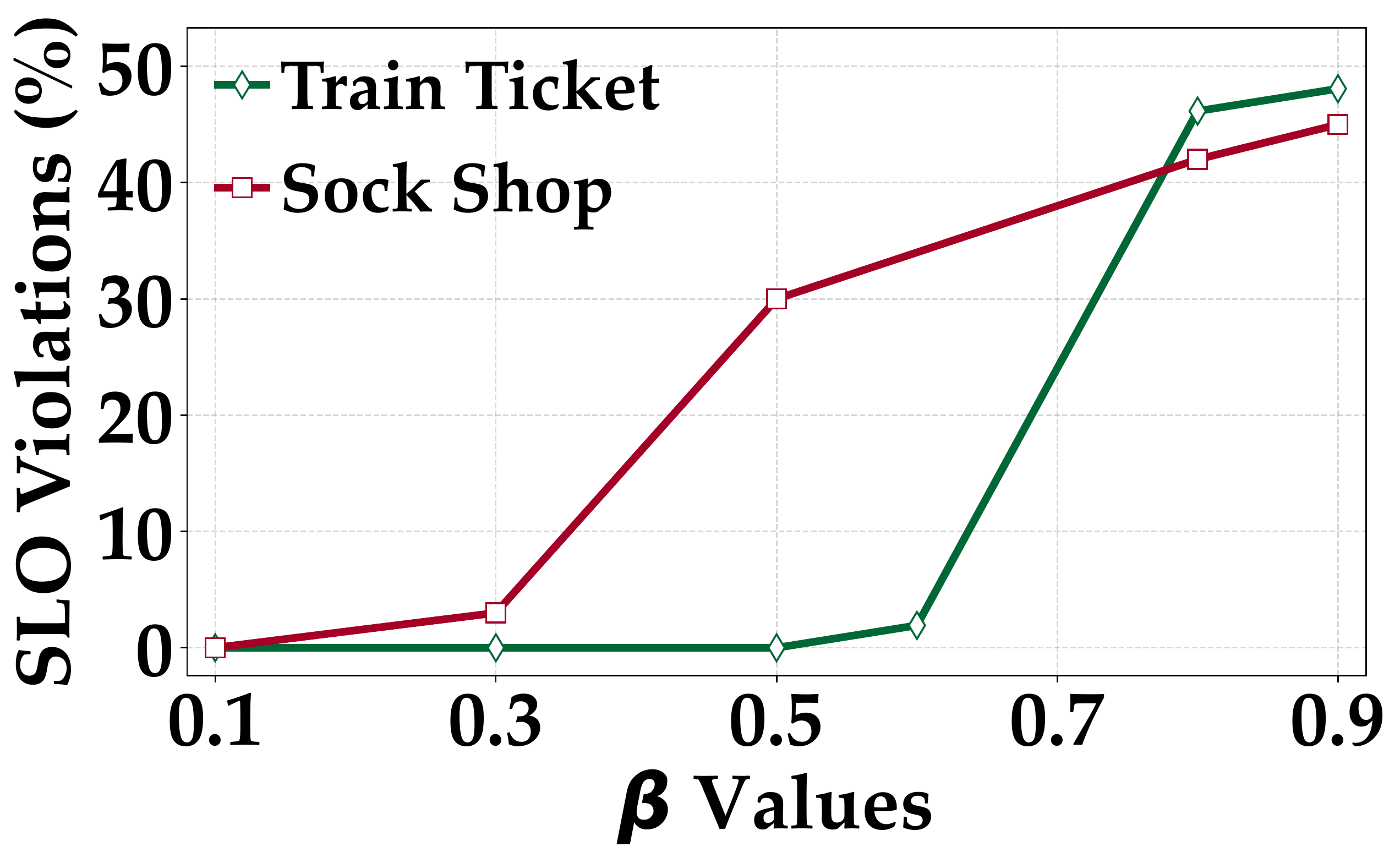}}
\caption{\ouralg's sensitivity to $\beta$ for a $\alpha=0.5$ (a) Resource allocation normalized to optimum. (b) SLO violations.}
\end{figure}

\subsection{Parameter Sensitivity}\label{sec:parametr_sensitivity}

Here we study how the two parameters $\alpha$ and $\beta$ affect \ouralg. Recall that $\alpha$ in Eqn.~\ref{eq:ns_equation} determines how aggressively we reduce resource - smaller $\alpha$ makes \ouralg reduce more resource for the same difference between response time and SLO. $\beta$, on the other hand, determines the maximum percentage resource reduction in each resource update iteration - smaller $\beta$ results in smaller resource change and vice versa. For this study, we run experiments on \trainT and \socS with workload 225 and 700 requests per second.

In Fig.~\ref{fig:combine_variable_alpha_fixed_beta_resource}, we show the change in resource allocation and in Fig.~\ref{fig:combine_variable_alpha_fixed_beta_slo_violations}, we show the number of SLO violations as we change $\alpha$. During this experiment, we keep $\beta=0.3$. We see that both smaller and larger values of $\alpha$ result in sub-optimal resource allocations for \trainT and \socS. This is because, for small $\alpha$, \ouralg is too aggressive making many SLO violations (as seen in Fig.~\ref{fig:combine_variable_alpha_fixed_beta_slo_violations}) and force to revert back to inefficient allocations. For high $\alpha$, on the other hand, \ouralg is slowed down prematurely at inefficient allocations, although it suffers much fewer SLO violations.

Next, in Figs.~\ref{fig:beta_CPU} and \ref{fig:beta_SLO}, we show the impact of change in $\beta$ while we keep $\alpha = 0.5$. Similar to our observation for $\alpha$ we see that aggressive resource reduction due to higher values of $\beta$ results in sub-optimal resource allocation while also suffering from many SLO violations.
While \ouralg is somewhat sensitive to both $\alpha$ and $\beta$, we can set $\alpha$ and $\beta$ for any system by tuning based on SLO violation. We can take a conservative approach, start with large $\alpha$ and small $\beta$, and gradually change their values keeping a close eye on the SLO violations.

\begin{figure}[t!]
\subfigure[]{\label{fig:burst_cpu}\includegraphics[width=0.23\textwidth]{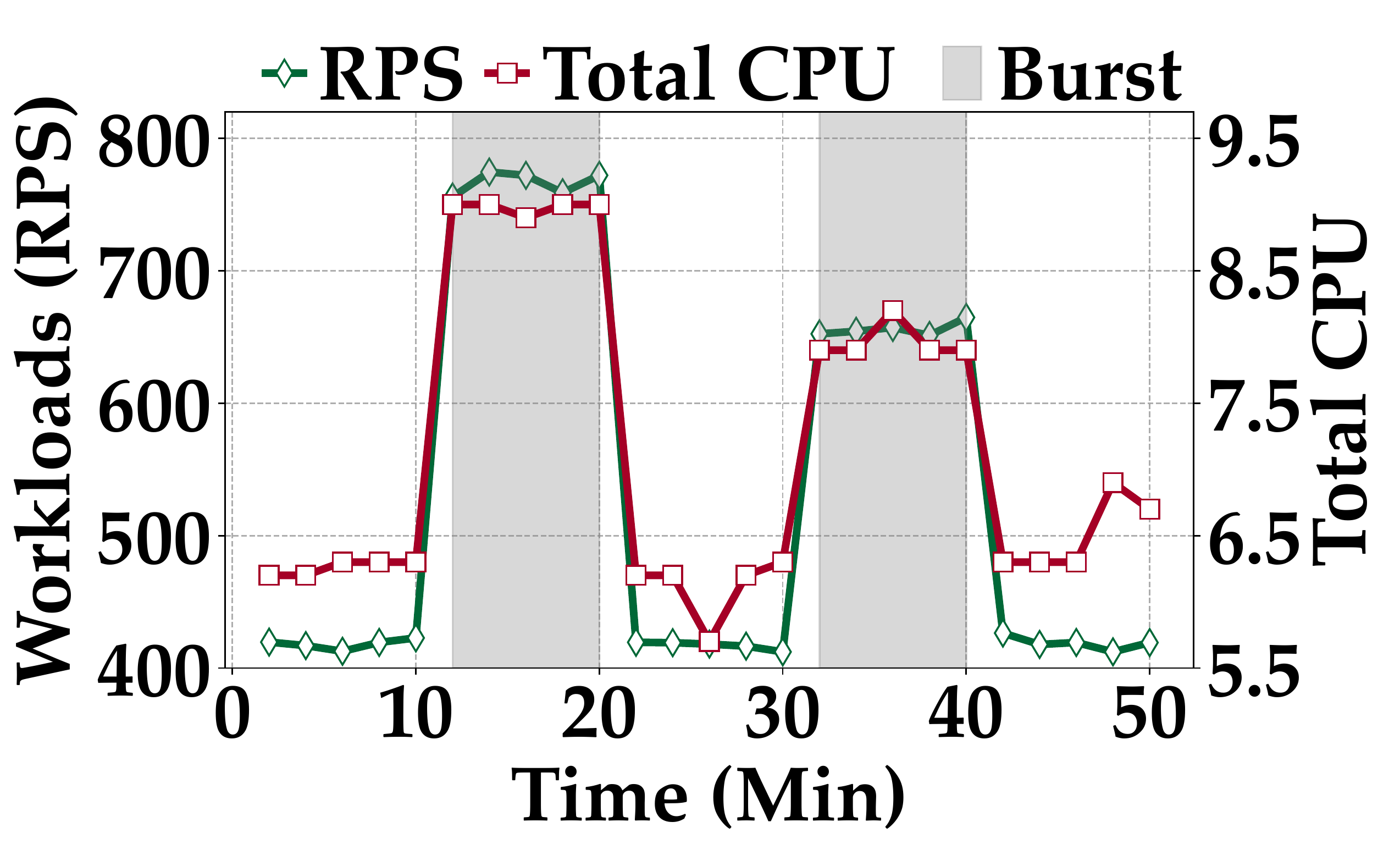}}
\subfigure[]{\label{fig:burst_response}\includegraphics[width=0.23\textwidth]{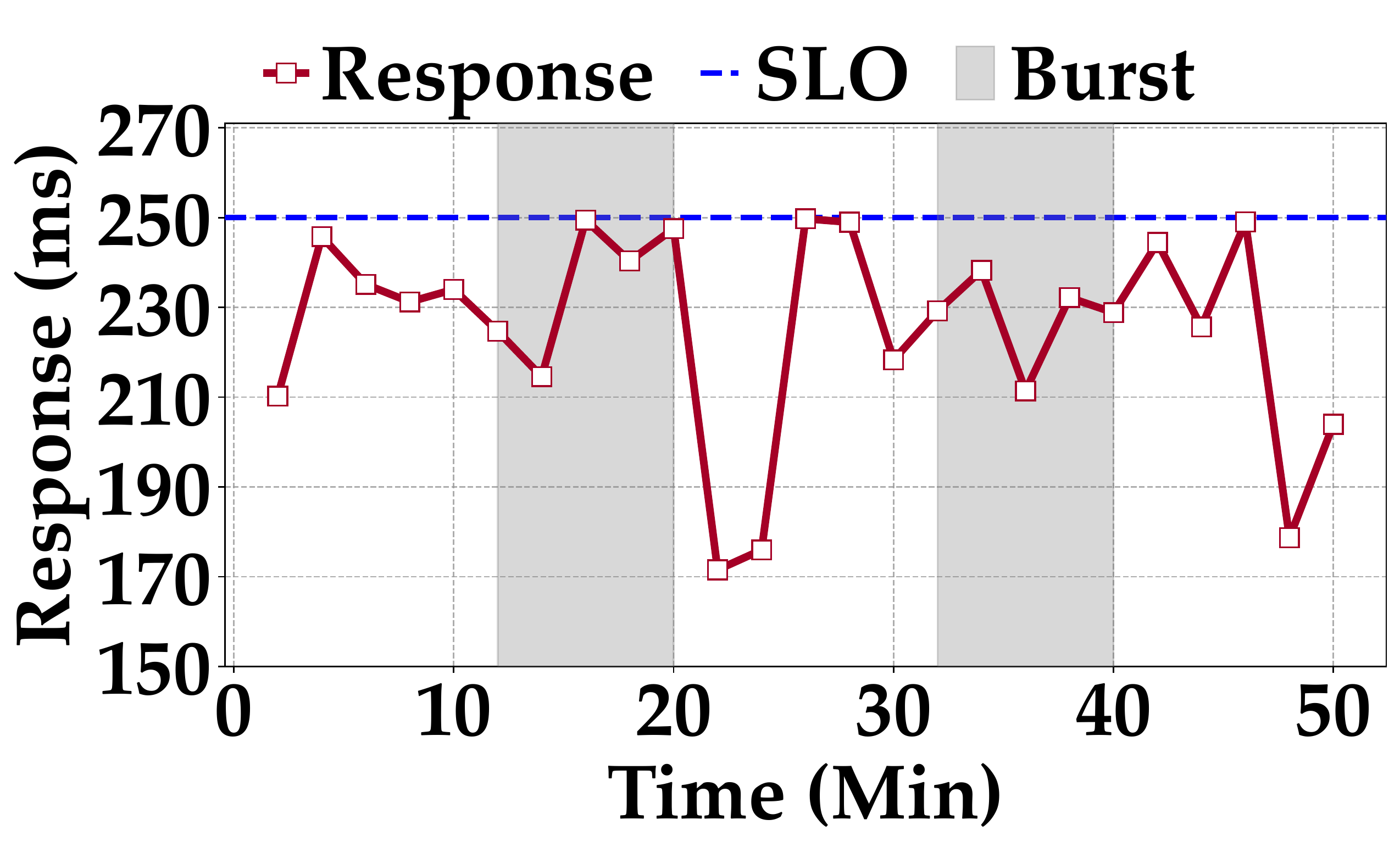}}
\caption{Operation of \ouralg with bursty workload in \socS. (a) Workload and CPU allocation. (b) Response time.}
\label{fig:ss_burst_test}
\end{figure}

\subsection{Adaptability}

\camera{
\textbf{Workload bursts.}
\ouralg can seamlessly handle sudden changes in workload. In Fig.~\ref{fig:ss_burst_test}, we show how \ouralg handles workload bursts for \socS by switching the resource allocation to the workload range corresponding to the workload burst. Here, we consider \ouralg has already traversed through the resource reduction iterations for all workload ranges. As shown in Fig.~\ref{fig:burst_cpu}, we create two workload burst of 10 minutes where the workload shoots up from 400 RPS to around 750 RPS and 650 RPS. We see that \ouralg quickly changes the CPU allocation to keep the response time below SLO (in Fig.~\ref{fig:burst_response}). Note here that, since we update the resource allocation every two minutes, \ouralg can react to a workload burst lasting less than two minutes. Nevertheless, we can adapt \ouralg to respond to short-lived workload bursts by reducing the resource update interval.}

\begin{figure}[t!]
\centerline{\includegraphics[width=0.5\textwidth]{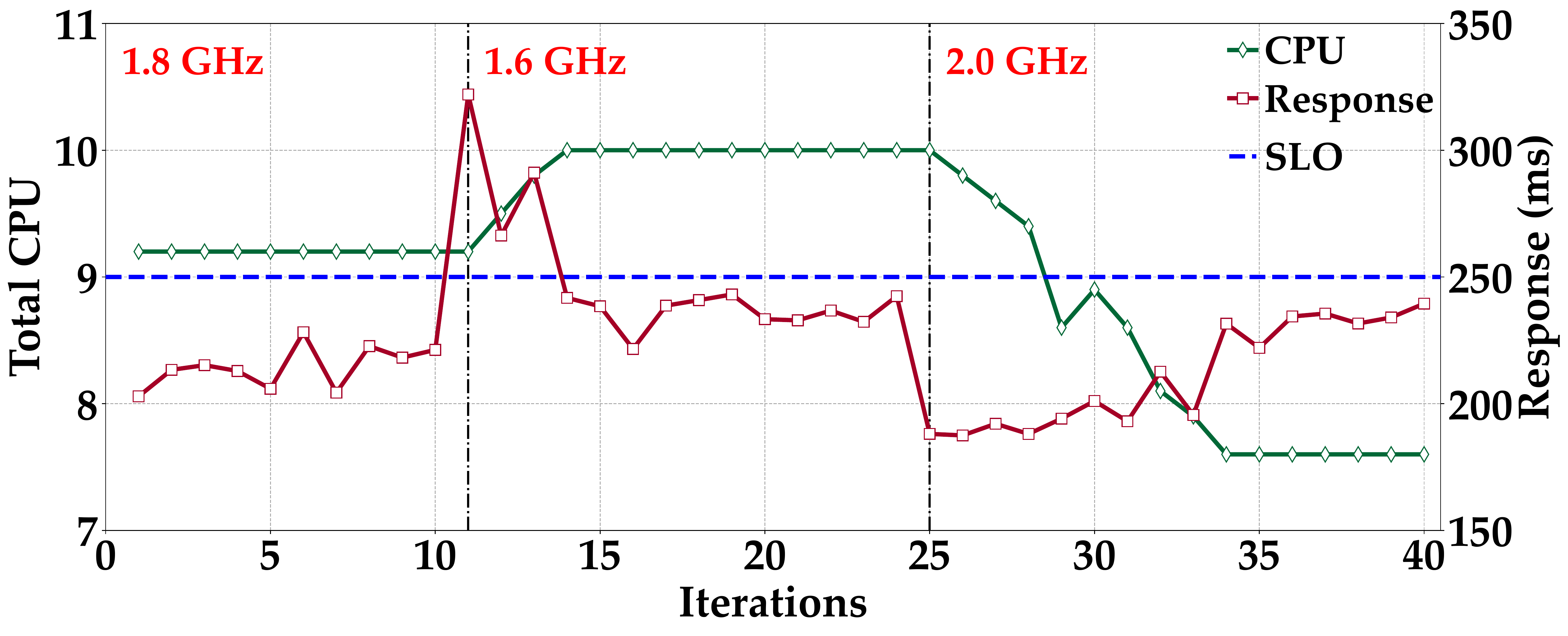}}
\caption{Adaptability of \ouralg to changes in CPU speed for \socS. The CPU speed change represents hardware or software updates that alters the resource demand.}
\label{fig:cpu_frequency_test}
\end{figure}

\textbf{Operating environment.}
Our \ouralg's lightweight design enables adaptability to operation condition changes. Such changes may lead to different response times even when the resource allocation is not altered. We change our server's CPU clock speeds from 1.8 GHz to 1.6 GHz and 2 GHz. These changes mimic a real-world scenario where a hardware or software change in the microservice alters the resource allocation dynamics.
While we make the clock speed changes, we use \ouralg to manage \socS's resource. A change in CPU frequency essentially changes the resource requirement for satisfying the SLO. Fig.~\ref{fig:cpu_frequency_test} shows the CPU allocation and the corresponding response time as we change the CPU frequency. We see that \ouralg can successfully change the resource allocation to satisfy the SLO demonstrating its capabilities to adapt.

\begin{figure}[t!]
\centerline{\includegraphics[width=0.5\textwidth]{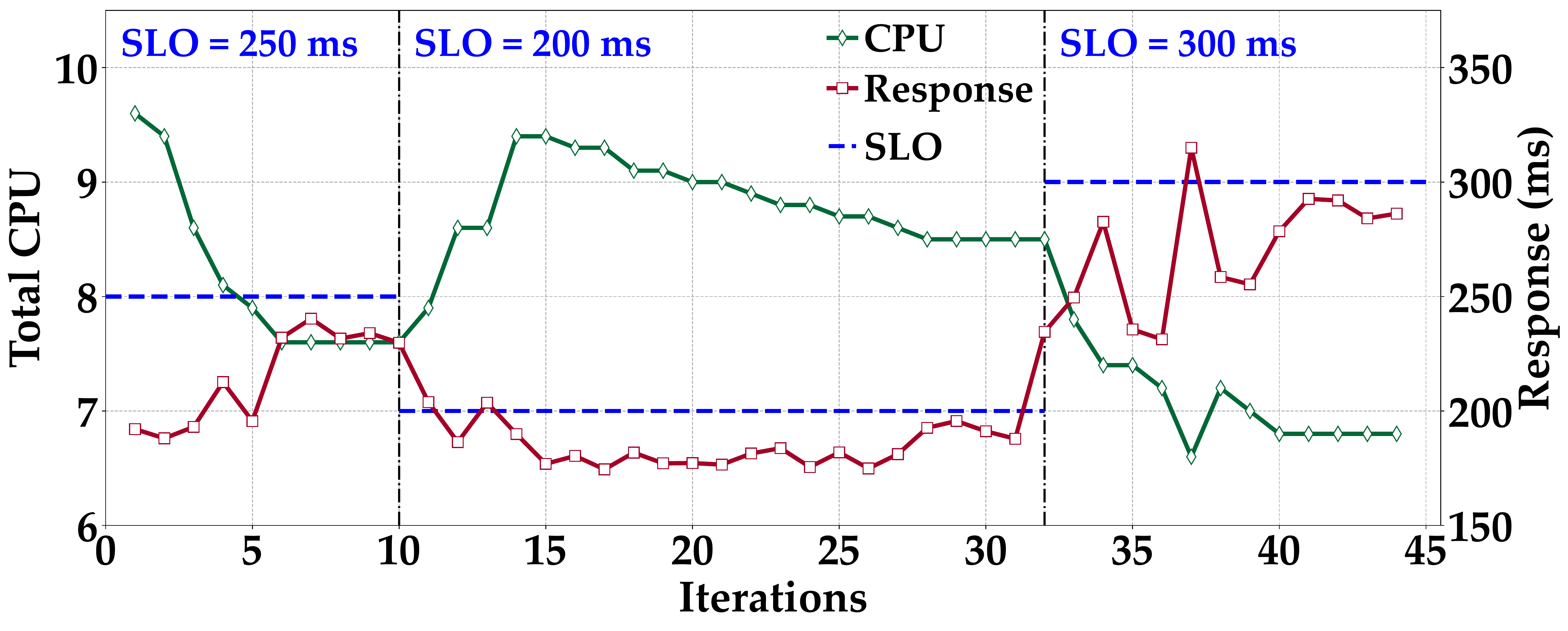}}
\caption{Adaptability of \ouralg to changes in SLO for \socS. Dynamic SLO can be used to trade performance for resource savings.}
\label{fig:cpu_SLO_change}
\end{figure}

\textbf{Dynamic SLO change.}
In Fig.~\ref{fig:cpu_SLO_change}, we show that \ouralg can also navigate towards efficient resource allocation as we change the SLO. Dynamically changing SLO can be a useful approach for applications that are willing to trade performance for resource savings to meet long-term goals such as cost budget \cite{islam2015online}. Dynamic SLO essentially adds another control knob for managing the microservices application.
Unlike existing ML-based microservice managements, which will need to retrain with new SLO,  \ouralg can quickly adapt to SLO changes and tune the resource accordingly.

\section{Related Works}
\textbf{Microservice autoscaling.} Resource autoscaling has been extensively studied in the public cloud domain \cite{autoscale_web_apps, gandhi2012autoscale,baarzi2019burscale,kalavri2018three,Wajahat2019MLscaleAM}. The recent advancement of microservices has attracted a similar interest in autoscaling of microservice-based applications in academic settings \cite{gias2019atom,rossi2020hierarchical}, as well as industrial settings \cite{kubernetes_autoscaler,google_autoscale}. These autoscalers implement rule-based approaches in resource management. For example, Kubernetes \cite{kubernetes_autoscaler} uses 90-th percentile resource usage in recent samples to set CPU and memory allocations with a 15\% overprovisioning. Google Autopilot \cite{google_autoscale} uses 95-th percentile for CPU and maximum for memory in the recent samples as a marker for resource allocation in the upcoming interval. Alternative to the rule-based approach, Google also uses ML-based autoscaling using a combination of reinforcement learning and time series analysis \cite{autopilot}.
\cite{kwan2019hyscale} also proposes rule-based autoscaling based on CPU and memory utilization. However, rule-based autoscaling requires deep application knowledge to set up the thresholds that can vary with application. Meanwhile, \cite{gias2019atom} proposes hybrid autoscaling based on analytical modeling using a layered queue network.

SHOWAR \cite{baarzi2021showar}, in spirit, is the closest to our design approach. It uses the variance in historical usage for vertical scaling and a proportional-integral-derivative (PID) controller for horizontal scaling. Nonetheless, SHOWAR still requires extensive tracing from the CPU scheduler for its scaling decision. \camera{On the other hand, similar to our opportunistic resource reduction, \cite{sharma2019resource} utilizes ``resource deflation'' where preemptible virtual machines' resources are dynamically controlled. However, while resource deflation gives away transient resources to avoid preemption, we use resource reduction as a mean to find efficient allocation by carving redundant resources.}

\textbf{SLO oriented resource management.}
In another line of work, ML-based approaches are used to identify and mitigate root causes of SLO violations in microservices \cite{zhang2021sinan,firm,gan2021sage,gan2019seer,hou2021alphar}. For example, Sinan \cite{zhang2021sinan} uses a neural network to estimate short-term performance and a boosted trees model to estimate long-term performance to make per tier resource allocation. Sinan allows SLO violations to identify corner cases for resource allocation. Seer \cite{gan2019seer} requires fine-grained tracing for building its model and SLO violating cases to train its deep neural network to identify QoS violations. AlphaR \cite{hou2021alphar}, on the other hand, uses neural graph networks to capture the complex relationship between microservices and estimate application performance for resource allocation. Despite their impressive results in capturing minute details of microservices, they heavily depend on data and are slow to dynamically changing conditions for microservices. In designing \ouralg, we depart from using complicated ML models and instead trade capturing microservice details for agility and adaptability in resource management.

\section{Concluding Remarks}\label{sec:conclusion}

In this paper, we proposed \ouralg, an iterative feedback-based approach to autoscaling microservices. \ouralg is lightweight as it only requires the applications end-to-end performance and microservice-level CPU utilization and CPU throttling to navigate to efficient microservice resource allocation. Utilizing the lightweight design, we also developed a novel approach of dynamic workload-ranging to make workload-aware resource allocation with \ouralg. Using three prototype microservice implementations, we showed that \ouralg can achieve a performance close to the optimum resource allocation and save as much as 33\% resource compared to commercially used rule-based resource allocation.

\textbf{Limitations of \ouralg's current implementation.} \ouralg's implementation has several limitations that we plan to address in its future iterations. First, when \ouralg causes an unintentional SLO violation, it rolls back the resource configuration in the next time step. Hence, the application suffers from bad performance during the entire resource update interval (e.g., 10 minutes). \ouralg can be improved by implementing higher resolution performance monitoring (e.g., within 10 seconds), catching the SLO violations early, and rolling back configuration to mitigate it.
Further, \ouralg rolls back the configuration to the most recent configuration without SLO violation. It does not take into account the degree of SLO violation. For instance, a QoS violation where the response time is significantly higher than the SLO indicates that \ouralg should roll back the configuration farther into the past to allocate more resources. On the other hand, while \ouralg logs the resource allocation of all microservices and response times in its allocation history database, RHDb, for rollback and exploration purposes, it does not utilize this information in its decision. Finally, \ouralg in this study only considers CPU resource allocation meanwhile memory and I/O resources allocation can also be important for microservices' performance depending on the nature of the application. Moreover, \ouralg also does not explicitly address the impacts and trade-offs among vertical (i.e increasing resource in one node) and horizontal (i.e., increasing the number of nodes) resource scaling.

\section{Acknowledgments}
This work is supported in parts by the US National Science Foundation under grant number CNS-2104925.

\bibliographystyle{ieeetr}
\balance
\bibliography{bibliography/main}

\end{document}